\documentclass[aoas]{imsart}

\RequirePackage{amsthm,amsmath,amsfonts,amssymb}
\RequirePackage[authoryear]{natbib}
\RequirePackage[colorlinks,citecolor=blue,urlcolor=blue]{hyperref}
\RequirePackage{graphicx}

\usepackage{multicol}
\usepackage{multirow}

\usepackage{placeins}

\newcommand{\bs}{\textbf{s}}

\begin{document}

\begin{frontmatter}
\title{Assessing marine mammal abundance: a novel data fusion}
\runtitle{Data fusion for marine mammal abundance}
\begin{aug}
\author[A]{\fnms{Erin M.} \snm{Schliep}\ead[label=e1]{emschliep@ncsu.edu}},
\author[B]{\fnms{Alan E.} \snm{Gelfand}\ead[label=e2]{alan@duke.edu}}
\author[C]{\fnms{Christopher W.} \snm{Clark}\ead[label=e3]{cwc2@cornell.edu}}
\author[D]{\fnms{Charles M.} \snm{Mayo}\ead[label=e4]{c.mayoiii@comcast.net}}
\author[D]{\fnms{Brigid} \snm{McKenna}\ead[label=e5]{bmckenna@coastalstudies.org}}
\author[E]{\fnms{Susan E.} \snm{Parks}\ead[label=e6]{sparks@syr.edu}}
\author[F]{\fnms{Tina M.} \snm{Yack}\ead[label=e7]{tina.yack@duke.edu}}
\and
\author[F]{\fnms{Robert S.} \snm{Schick}\ead[label=e8]{robert.schick@duke.edu}}
\address[A]{Department of Statistics, North Carolina State University \printead{e1}}

\address[B]{Department of Statistical Science, Duke University \printead{e2}}

\address[C]{K. Lisa Yang Center for Conservation Bioacoustics, Cornell Lab of Ornithology, Cornell University \printead{e3}}

\address[D]{Right Whale Ecology Program, Center for Coastal Studies \printead{e4,e5}}

\address[E]{Biology Department, Syracuse University \printead{e6}}

\address[F]{Nicholas School of the Environment, Duke University \printead{e7,e8}}
\end{aug}

\maketitle

\begin{abstract}

Marine mammals are increasingly vulnerable to human disturbance and climate change. Their diving behavior leads to limited visual access during data collection, making studying the abundance and distribution of marine mammals challenging. In theory, using data from more than one observation modality should lead to better informed predictions of abundance and distribution. With focus on North Atlantic right whales, we consider the fusion of two data sources to inform about their abundance and distribution. The first source is aerial distance sampling which provides the spatial locations of whales detected in the region. The second source is passive acoustic monitoring (PAM), returning calls received at hydrophones placed on the ocean floor. Due to limited time on the surface and detection limitations arising from sampling effort, aerial distance sampling only provides a partial realization of locations. With PAM, we never observe numbers or locations of individuals. To address these challenges, we develop a novel \emph{thinned} point pattern data fusion. Our approach leads to improved inference regarding abundance and distribution of North Atlantic right whales throughout Cape Cod Bay, Massachusetts in the US.  We demonstrate performance gains of our approach compared to that from a single source through both simulation and real data.

\end{abstract}


\end{frontmatter}
\doublespacing
\section{Introduction}
As the field of marine mammal science progresses, our ability to collect \textit{more} data and data of disparate nature grows \citep{coutinho2021north, bucklandWildlifePopulationAssessment2023}.
Accordingly, it is imperative that we develop probabilistic approaches and inferential techniques to jointly use as much of the available data as possible to answer increasingly complex biological questions.
Because marine mammals are inherently difficult to study \citep{NOWACEK2016235}, enhancing our understanding of these biological processes often entails combining, henceforth \emph{fusing}, data arising from different approaches to data collection.

In the statistics literature, \emph{data fusion} is synonymous with the terms data assimilation, data aggregation, data melding, and data integration.
The shared objective underlying each of these approaches is to combine two or more datasets that are assumed to be informing about the same process of interest but that are different in terms of their source, type, uncertainty, and/or spatial or temporal resolution.
Examples of disparate data sources in environmental settings include monitoring station networks, satellites, and computer model outputs to inform about important climate variables or \emph{in situ} observations and remote sensing data to inform about species presence or abundance.
Together, these data can improve our understanding of the underlying process of interest.

Process-based data fusion is more than an approach to increase sample size.  First, it responds to the nature of the data collection for each different data type, considering stochastically the type of information that each provides about the underlying process. Second, probabilistic dependence between the sources is accounted for through the modeling structure. Finally, it can appropriately assess and account for the uncertainty associated with estimating the process features and in predicting the process behavior.
Bayesian hierarchical models provide a natural scaffolding for such process-based approaches.

Our process-based data fusion is directed towards two types of marine mammal data---visual data from plane-based distance sampling and acoustic data from a network of passive acoustic recorders.
Distance sampling \citep{bucklandIntroductionDistanceSampling2001,bucklandAdvancedDistanceSampling2007,bucklandDistanceSamplingMethods2015}
and density surface modeling \citep{hedleySpatialModelsLine2004,millerSpatialModelsDistance2013,yuanPointProcessModels2017} is perhaps the most prominent way to generate spatial estimates of density for marine mammals. The data are obtained from planes or ships that are conducting transects where the $x,y$ position of detected animals is recorded.
These $x,y$ positions of detected animals and their perpendicular distance from the trackline are used to estimate detection functions.
The relationship between the locations and the environment are often estimated with generalized additive models, and density estimates are generated \citep{millerSpatialModelsDistance2013}.
Alternatively, the $x,y$ data can be treated as a spatial point pattern and used to infer about abundance as in \cite{Johnson2010} and \cite{yuanPointProcessModels2017}.

Ecologists have also used acoustics data to estimate the density of marine species \citep{marquesEstimatingNorthPacific2011, marquesEstimatingAnimalPopulation2013}. 
Conceptually, the location of the monitor is analogous to points within a point transect. 
Similar approaches have been used for estimating avian populations using sightings along driven transects \citep{bucklandPOINTTRANSECTSURVEYSSONGBIRDS2006}. 
With acoustic data, distances are not observed and have to be inferred based on principles of underwater sound propagation, and the acoustic and behavioral characteristics of the animal. With time difference of arrival information, i.e., information about when the same sound was detected on multiple hydrophones, locations can be estimated with error \citep{watkinsSoundSourceLocation1972a}.

\begin{figure}
    \centering
\includegraphics[scale=.35]{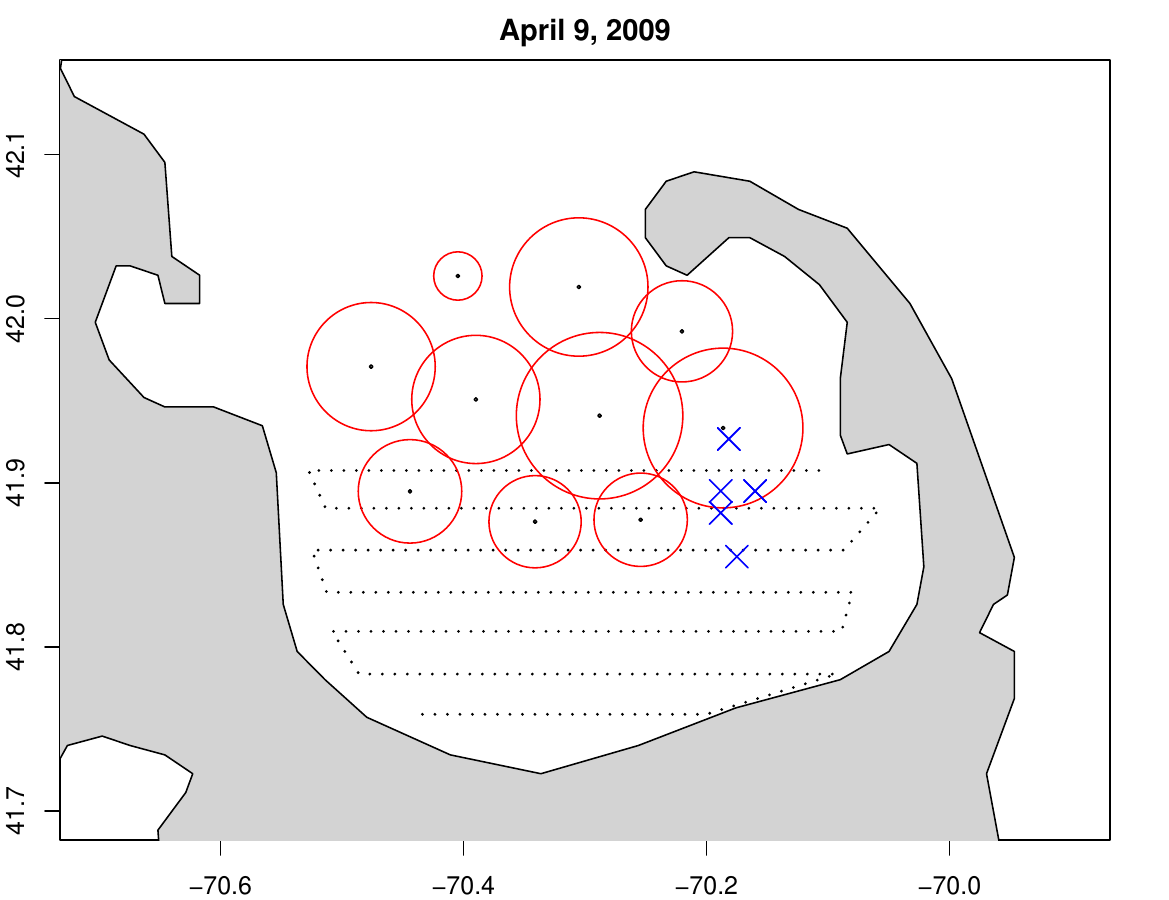}
\includegraphics[scale=.35]{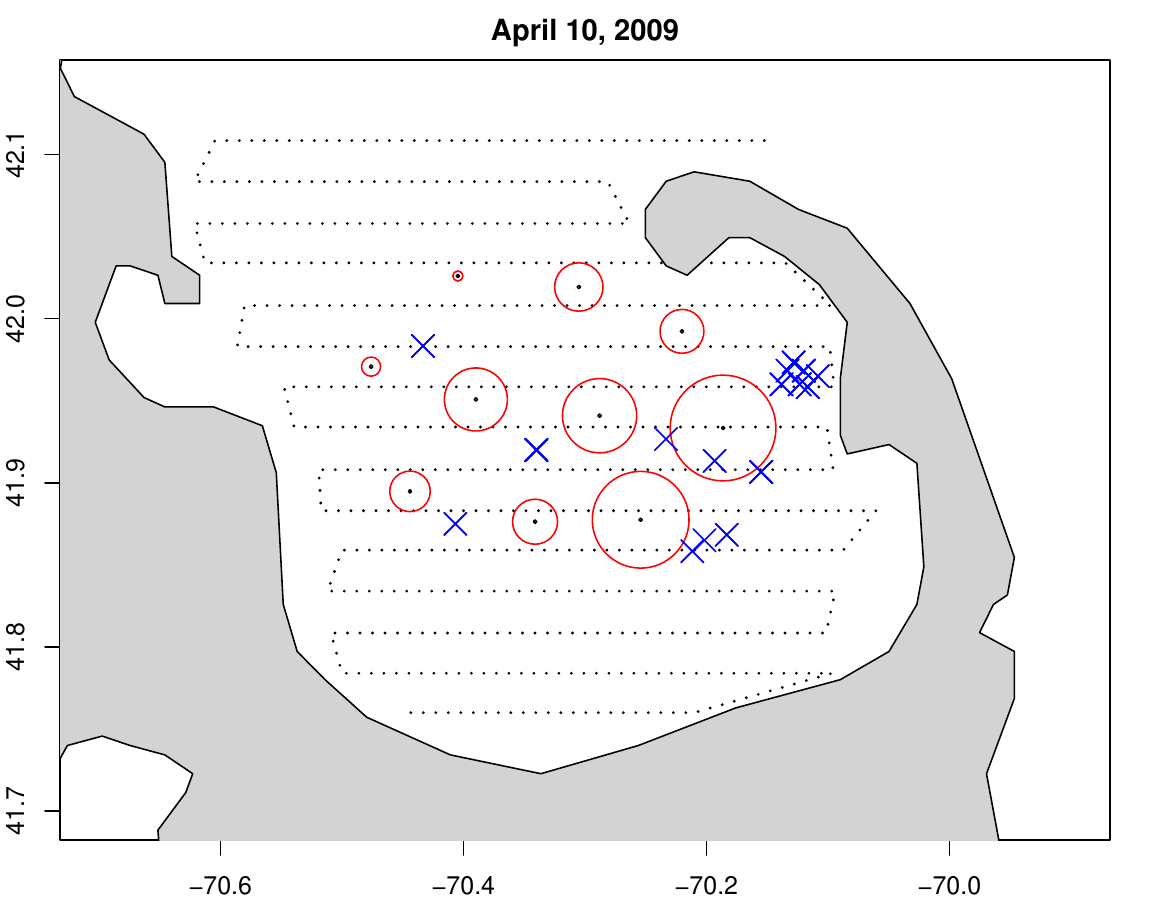}
    \caption{Data collection for Cape Cod Bay on April 9, 2009 (left) and April 10, 2009 (right) using aerial transects (dotted lines) and passive acoustic recorders. Aerial detections are depicted as {\color{blue} $\times$} and the number of calls heard on each PAM are shown proportionally by the size of {\color{red} $\circ$} centered on the PAM location.}
    \label{fig:CCBdata}
\end{figure}

Our analysis addresses questions pertaining to the distribution and abundance of the North Atlantic right whale (NARW) population in Cape Cod Bay (CCB), Massachusetts. Aerial data for monitoring NARW populations have been collected by the Center for Coastal Studies in Provincetown, MA since 1998 \citep{mayoDistributionDemographyBehavior2018}. Aerial surveys follow a series of 15 east-west tracklines in CCB, which are separated by 2.8 km. These tracklines are flown from January through May, corresponding with the seasonal habitat of right whale. 
Peak numbers of individual NARW occur in late April \citep{mayoDistributionDemographyBehavior2018}. During these surveys, the goal is to detect, count, and individually identify NARWs \citep{clarkVisualAcousticSurveys2010a}. In the early 2000's, a passive acoustic monitoring (PAM) program was established, whose deployment was timed to coincide with the aerial surveys \citep{clarkVisualAcousticSurveys2010a}. 
The PAM effort consisted of an array of bottom-mounted hydrophones (known as Marine Autonomous Recording Unit---MARUs), continuously deployed in a variety of spatial configurations \citep{clarkVisualAcousticSurveys2010a}. 
These record ambient and bioacoustic data, from which a time series of right whale up-calls \citep{clarkAcousticRepertoireSouthern1982} can be annotated. These time-stamped up-calls form the basis of the acoustic data analyzed here. 
Figure \ref{fig:CCBdata} shows the data collected in Cape Cod Bay on April 9 and 10, 2009 for these two data sources. 
The tracklines flown show that only half of the region was surveyed while on April 9 while on April 10 there was a complete survey of CCB.
Aerial detections are depicted as {\color{blue} $\times$} and the number of calls heard on each PAM are shown proportionally by the size of {\color{red} $\circ$} centered on the PAM location.

Neither aerial distance sampling nor acoustic data on marine mammals collected in CCB are adequate to address questions pertaining to abundance and distribution due to their limited spatial and temporal coverage, and missed or mis-classified detections associated with the data collection processes.
As such, data fusion is a promising alternative as it leverages both data sources simultaneously.
Visual and acoustic data have been examined jointly in the marine mammal literature in spatially implicit ways to estimate abundance of migrating bowhead whales \citep{raftery1990bayes,fristrupCombiningVisualAcoustic1997}. However, there are few examples of fusing both visual and acoustic data to better estimate spatially explicit abundance (though see \cite{frasierCetaceanDistributionModels2021} or \cite{sigourney2023}).

The most common approaches for spatial and spatio-temporal data fusion modeling for environmental and ecological data are either geostatistical data fusion or the fusion of geostatistical with areal data.
For example, fusion of monitoring station networks and computer model output has been developed for studying air pollution \citep{fuentesModelEvaluationSpatial2005, Rundel2015}, ozone \citep{zidek2012combining, liu2011empirical}, and agricultural crop yields \citep{finley2011improving}.
A multivariate data fusion model was developed by \cite{Schliep2020} for speciated lake nitrogen observations obtained for multiple species compositions using varying methods of reporting.
\cite{Wikle2001} and \cite{Wikle2005} develop a data fusion model for wind, including stream function, using satellite observations and weather center computer model output.
\cite{foley2008} present a fusion model for wind field data collected by monitors on surface buoys, ships, aircrafts, and coastal platforms as well as numerical ocean models. Data fusion of aerosols using multiple remote sensing instruments has also been proposed \citep{Nguyen2012}.

Our focus is on the fusion of two data sources which are imagined through point patterns.
In fusing point patterns, the most natural assumption would be that each source provides a full realization of a point pattern where the focus of the modeling is to learn about the nature of the intensity that is driving each realization. In such a case, a common intensity may be assumed for the two point patterns, e.g., a log Gaussian Cox process (LGCP) \citep[]{illian2008statistical}.  Both point patterns are viewed as conditionally independent given this intensity, resulting in a product form for the two point pattern likelihoods.  The model fitting becomes analogous to that for fitting an LGCP model to a single data source \citep{illian2008statistical,gelfand2018bayesian}.

The fusion required in our setting assumes that there is a single \emph{true} point pattern which yields our observed point pattern data, each of which arises as a partial realization of the true point pattern.
In model fitting, we adopt a model for the intensity surface from which the point pattern realization arises. The two data sources emerge as thinned process realizations.
We present approaches, which arise exclusively through hierarchical latent process modeling, that are generative -- the proposed model specification could produce the data that have been observed from each source.
Model fitting within the Bayesian framework enables full inference with regard to estimation, prediction, and uncertainty assessment.

Using both a broad simulation study and analysis of real data, we show better predictive accuracy and precision in terms of the distribution and abundance of marine mammals. Combining data sources also leads to greater spatial coverage which enables interpolation, rather than requiring potentially risky extrapolation \citep{yuanPointProcessModels2017}.
The simulation study allows us to illuminate the extent of the benefit across different sampling scenarios. This can aid in developing future sampling designs for aerial distance sampling data collection and placement of passive acoustic recorders.


The format of the paper is as follows. 
In Section \ref{sec2}, we introduce the novel fusion of two point pattern data sources and metrics for model assessment and comparison. 
Section \ref{sec3} provides a detailed simulation of the fusion modeling to demonstrate its benefit over a range of scenarios and to inform sampling designs for future data collection.  
Section \ref{sec4} turns to the fusion for real data collected on NARW for two different days over CCB.  Section \ref{sec5} concludes with a brief summary and further ongoing work.

\section{Data fusion for point pattern data}
\label{sec2}
\subsection{Introduction to point pattern data fusion}
\label{sec21}

We assume that there is a single true unobservable point pattern and each data source provides only partial realization of the full point pattern.  The realizations are \emph{partial} in the sense that, due to sampling effort or detection, only a portion of the true point pattern was seen.  We assume a sampling model for the true intensity, $\lambda_{true}(\bs)$ which has generated a point pattern realization, $\mathcal{S}$ of size $N$ over $\mathcal{D}$.  Under the most basic setting for two sources, we would observe $\mathcal{S}_{1}= \{\bs_1,...,\bs_{N_{1}}\}$ and $\mathcal{S}_{2} = \{\bs_{1},...,\bs_{N_{2}}\}$, each a \emph{thinned} version \citep{illian2008statistical,gelfand2018bayesian} of $\mathcal{S}$ where $N_1, N_2 \leq N$.
The observed point pattern data, $\mathcal{S}_{1}$ and $\mathcal{S}_{2}$, are the result of thinning mechanisms, $p_{1}(\bs)$ and $p_{2}(\bs)$, respectively, applied independently to the latent realization $\mathcal{S}$ from $\lambda_{true}(\bs)$.  This is the \emph{sampling} model and the resulting observed point pattern sources are conditionally independent given the thinning mechanisms and the latent point pattern.

The \emph{fitting} model specification employs both of the thinning mechanisms along with a choice of intensity model.  The thinning functions are specified to capture sampling effort, location availability, and detection (or misclassification). For example, assume a LGCP model for the intensity $\lambda(\bs)$.  The resulting intensities in the fitting model become $p_{1}(\bs)\lambda(\bs)$ and $p_{2}(\bs)\lambda(\bs)$.
See \cite{chakrabortyPointPatternModelling2011} for an illustration of this version of thinning.

Evidently, $\lambda(\bs)$ is not the true intensity, $\lambda_{true}(\bs)$.
That is, the fitting model is not the same as the generating, or sampling, model.  
The fitting model treats the observed thinned data as conditionally independent given the thinning mechanisms and the intensity. As such, the full likelihood is written
\begin{equation}
    \mathcal{L}(\lambda(\bs), \bs \in \mathcal{D}; \{\mathcal{S}_1\}, \{\mathcal{S}_2\}) = e^{-\lambda_1(\mathcal{D}) -\lambda_2(\mathcal{D})} \prod_{\bs_{i} \in \mathcal{S}_1}p_{1}(\bs_{i})\lambda(\bs_{i})
      \prod_{\bs_{j} \in \mathcal{S}_2}p_{2}(\bs_{j})\lambda(\bs_{j})
     \label{eq:liksource0}
\end{equation}
where $\lambda_1(\mathcal{D}) =\int_{\mathcal{D}} p_1(\bs)\lambda(\bs)d\bs$ and $\lambda_2(\mathcal{D}) =\int_{\mathcal{D}} p_2(\bs)\lambda(\bs)d\bs$.
Employing both thinning sources, the primary inference objective is to learn about $\lambda(\bs)$.
This intensity enables inference regarding the spatial variation of the point pattern process as well as expected counts in subsets of the region.  If there are parameters in the thinning mechanisms, additional scientific information will be needed in order to identify both the intensity and the thinning function. Alternatively, knowledge regarding the total expected number over the region could be used to inform the thinning mechanisms. Further details pertaining to model identifiability are given across Sections \ref{sec3} and \ref{sec4}.

One important benefit of the data fusion model over either of the single source models is the reduction in uncertainty. To illustrate, suppose we adopt a constant thinning for each data source such that $p_1(\bs)=p_1$ and $p_2(\bs)=p_2.$ Then, $\mathcal{S}_{1}$ is a realization over the region $\mathcal{D}$ from an NHPP (or LGCP) with intensity $p_{1}\lambda(\bs)$ and $\mathcal{S}_{2}$ is a realization from an NHPP (or LGCP) with intensity $p_{2}\lambda(\bs)$. 
Let $N_1$ be the random number of points in $\mathcal{S}_{1}$ and $N_2$ be the random number of points in $\mathcal{S}_{2}$.  Assume the goal is to estimate the expected number of points, $N$, in $\mathcal{S}$ in $\mathcal{D}$, computed as $E(N) = \lambda(\mathcal{D}) = \int_{\mathcal{D}} \lambda(\bs)d\bs$ where $\mathcal{S}$ is a realization from an NHPP (or LGCP) with intensity $\lambda(\bs)$.  Then, since $E(N_1) = p_1 \lambda(\mathcal{D})$ and $E(N_2) = p_2 \lambda(\mathcal{D})$, $N_1/p_1$ and $N_2/p_{2}$ are both unbiased estimators of $\lambda(\mathcal{D})$. If $p_1 < p_2$, the variance of the latter is smaller than the variance of the former, which aligns with our intuition that we can better estimate $\lambda(\mathcal{D})$ with less thinning. Under data fusion, the joint thinning of $\mathcal{S}$ yields a thinning probability $1-(1-p_{1})(1-p_{2}) > \text{max}(p_{1}, p_{2})$.  Data fusion results in less thinning than either source, meaning we can expect to better estimate $\lambda(\mathcal{D})$ and better predict $\mathcal{S}$.  

We can extend this argument to the case of spatially varying thinning, $p_{1}(\bs)$ and $p_{2}(\bs)$.  Suppose we partition $\mathcal{D}$ into arbitrarily fine disjoint sets $B_{k}$ such that if $\bs \in B_{k}$, $p_{1}(\bs) = p_{1k}$ and $p_{2}(\bs) = p_{2k}$ where $p_{1k} < p_{2k}$ $\forall k$.  We can consider the two unbiased estimates for $\lambda(\mathcal{D})$, $\sum_{k} N_{1k}/p_{1k}$ and $\sum_{k} N_{2k}//p_{2k}$, where $N_{ik}$ denotes the random number of points in $\mathcal{S}_i$ in the set $B_{k}$.  Since $N_{ik}$ and $N_{ik'}$ are independent for $k \neq k'$, $Var\left( \sum_{k} N_{1k}/p_{1k}\right) = \sum_{k} \lambda(B_{k})/p_{1k} > \sum_{k}\lambda(B_{k})/p_{2k} = Var \left(\sum_{k} N_{2k}/p_{2k}\right)$. 
Again, this implies that we can better estimate $\lambda(\mathcal{D})$ using the data fusion given the joint thinning of $\mathcal{S}$.

\subsection{Novel point pattern data fusion for NARW data}
The analysis for the NARW data is more complex than the basic setup above.
These data sources are inherently of different data type with different spatial and temporal scales as well as have varying sources of uncertainty and detection probabilities.
Aerial distance sampling \citep{mayoDistributionDemographyBehavior2018,ganleyWhatWeSee2019} provides a partially observed realization of the full point pattern for each transect. These data are collected roughly bi-weekly throughout the NARW season in CCB. The data exhibit degradation/thinning due to time on surface, sampling effort, and detection probability \citep{ganleyWhatWeSee2019}.
The PAM data source collects data continuously, however it does not provide observation locations. 
The PAM data provide right whale ``up-calls'' \citep{clarkAcousticRepertoireSouthern1982} at a set of hydrophone detectors within a network/platform of detectors \citep{clarkVisualAcousticSurveys2010a}. Given a specified time window (e.g., hourly), we obtain the number of calls associated with each hydrophone.  Adjusting for call rates \citep{parksSoundProductionBehavior2011} and an acoustic detection function \citep{palmerEvaluationCoastalAcoustic2022,palmerAccountingLombardEffect2022a}, the expected number of calls received by a monitor is viewed as an integral of the degraded intensity over the detection area.


Here, let $\mathcal{S}_{dist}$ and $\mathcal{S}_{PAM}$ denote the two sources with thinning mechanisms $p_{dist}(\bs)$ and $p_{PAM}(\bs)$, respectively. For the distance sampling data source, $\mathcal{S}_{dist}$ consists of $L$ point pattern realizations from $\mathcal{S}$, each corresponding to a distinct aerial transect.  For $\ell =1, \dots, L$, the realization $\mathcal{S}_{dist,\ell}$ arises as a result of applying independent Bernoulli trials with thinning mechanism, henceforth detection function, $p_{dist,\ell}(\bs)$, to all of the points $\bs \in \mathcal{S}$.  The set of trials resulting in a ``1'' provides the observed point pattern associated with transect $\ell$.
For the passive acoustic monitoring data source, $\mathcal{S}_{PAM}$ is observable in the form $Y_k$, $k=1, \dots, K$, where $Y_k$ is the number of calls detected at a point source located at $\mathbf{h}_{k}$ for $\mathbf{h}_{k} \in \mathcal{D}$; no actual point locations are observed.  The thinning mechanism, $p_{PAM,k}(\bs)$, $k=1, \dots, K$, arises from the detection capability of point source $k$.  Further details pertaining to $p_{dist,\ell}(\bs)$ and $p_{PAM,k}(\bs)$ are supplied in Section \ref{sec3}.

In terms of the modeling, the likelihood for the distance sampling will take the form
\begin{equation}
    \mathcal{L}(\lambda(\bs), \bs \in \mathcal{D}; \{\mathcal{S}_{dist,\ell}\}) = e^{-\sum_{\ell}\lambda_{dist,\ell}(\mathcal{D})} \prod_{\ell}\prod_{\bs_{i} \in \mathcal{S}_{dist,\ell}}p_{dist,\ell}(\bs_{i})\lambda(\bs_{i})
    \label{eq:liksource1}
    \end{equation}
where $\lambda_{dist,\ell}(\mathcal{D}) = \int_{\mathcal{D}} p_{dist,\ell}(\bs)\lambda(\bs)d\bs$
with $p_{dist,\ell}(\bs)$ the detection function associated with transect $\ell$.  This modeling is in the spirit of \cite{Johnson2010} and \cite{yuanPointProcessModels2017} who also view the observed aerial data as independent realizations of a point pattern for each transect route.
For the passive acoustic monitoring, the likelihood has the form 
\begin{equation}
    \mathcal{L}(\lambda(\bs), \bs \in \mathcal{D}; \{Y_k\}) = e^{-\sum_{k}\lambda_{PAM,k}} \prod_{k}\lambda_{PAM,k}^{Y_k}.
    \label{eq:liksource2}
    \end{equation}
Here,  $\lambda_{PAM,k}= \int_{\mathcal{D}}p_{PAM,k}(\bs)\lambda(\bs)d\bs$.

The observed data, $\mathcal{S}_{dist}$ and $\mathcal{S}_{PAM}$, are conditionally independent given $\lambda(\bs)$ and the thinning mechanisms, but are marginally dependent.
Therefore, the likelihood function for the data fusion model is the product of (\ref{eq:liksource1}) and (\ref{eq:liksource2}).
With $\lambda(\bs)$ assumed to be a LGCP, discretization is required for all of the stochastic integrals following \cite{gelfand2018bayesian}.
With an appropriate model for $\lambda(\bs)$,
the formal inference objectives are to learn about the spatial variation in the intensity as well as abundance globally (or locally) by integrating the intensity over the entire study region or subregions.

\subsection{Fusion performance}

To demonstrate the performance of the data fusion in our simulation study, we employ three metrics where we know $\lambda_{true}(\bs)$ and $\mathcal{S}$. These metrics can be used to assess performance on recovering the true point pattern, $\mathcal{S}$ and $N$, as well as the intensity surface and $\lambda(\mathcal{D})$.  First, we compare true abundance (total or for a subregion) with the posterior predictive distribution of abundance under the individual data source and data fusion models.  We can directly calculate the root mean square error (RMSE) to compare the posterior mean abundance with the truth.  With posterior predictive realizations, we can also compute the ranked probability score (RPS) \citep{gneiting2007strictly}.
Graphically, we can overlay the posterior predictive distributions from each of the models on the actual realization for a clear depiction of accuracy and precision.

A second metric is a version of a full data likelihood in the context of missing data.  That is, we can consider the posterior mean log$L(\lambda(\bs), \bs \in \mathcal{D}; \mathcal{S})$ under each model fitting.  Specifically, we are not calculating the log likelihood for each model given its data but rather the log likelihood associated with each model given we have the full point pattern.  Again, we have a posterior distribution for the log likelihood for each model to make probabilistic comparisons. 

A third metric compares $\lambda_{true}(\bs)$ with the estimated intensities, $\lambda(\bs)$, arising from fitting each of the models.  Two possible summary measure include $\frac{1}{|\mathcal{D}|}\int_{\mathcal{D}} |\hat{\lambda}(\bs) - \lambda_{true}(\bs)| d\bs$ or $\frac{1}{|\mathcal{D}|}\int_{\mathcal{D}} (\hat{\lambda}(\bs) - \lambda_{true}(\bs))^2 d\bs$ where $|\mathcal{D}|$ is the area of $\mathcal{D}$.
We can approximate the integral using a fine grid.

\section{A detailed simulation example}
\label{sec3}

The challenge in arguing on behalf of the data fusion is that we can never observe a realization of the entire NARW population in CCB at any given time.
With known simulated realizations of full, \emph{true} point patterns, we aim to demonstrate what benefits exist (and when) for estimating species abundance across a landscape through fusion of the two data sources.
To begin, we fully elaborate a particular simulation to reveal all aspects from specification to inference.  Then, we broaden our simulation by varying the features of the data sources and of the environment in order to learn how the performance of the fusion varies across a range of scenarios.
These simulations are designed to provide true data and sampled observations resembling that from the CCB study region using the knowledge of the sampling protocols for each data source.

For the simulation, we imagine a spatial domain over a sufficiently short window in time such that the true locations of all of the whales in the region can be viewed as a spatial point pattern over the region.
Importantly, it is not a spatio-temporal point process but rather a conceptual picture of a spatial point pattern at any time.
Lastly, since we lack whale depth data, the latent intensity surface is viewed conceptually as an intensity integrated over depth.


\subsection{True location simulation}

To simulate \emph{true} abundance along with \emph{observed} aerial and PAM data in CCB, we start with a spatially explicit intensity surface $\lambda_{true}(\mathbf{s})$ for $\mathbf{s} \in \mathcal{D}$.
We define an illustrative spatial domain $\mathcal{D} = [0,40] \times [0,40]$, where the units are kilometers (km). This area roughly approximates the area of CCB. 
For the intensity function, we assume a realization of a LGCP arising through a mixture of three Gaussian processes, viewed as covariates.
Specifically, we define the true intensity surface as
$$\lambda_{true}(\mathbf{s}) = \exp{\beta_0 + \beta_1 X_1(\mathbf{s}) + \beta_2 X_2(\mathbf{s})+ \beta_3 X_3(\mathbf{s})}$$
where $\beta_0$, $\beta_1$, $\beta_2$ and $\beta_3$ are coefficients and $X_1(\mathbf{s})$, $X_2(\mathbf{s})$, and $X_3(\mathbf{s})$ become spatial covariates. For each covariate, we discretize the region using a 0.5 $\times$ 0.5 resolution grid and generate a realization from a mean 0 GP at the collection of grid cell centroids. We assume an exponential covariance function with an effective range of 9 to capture local spatial dependence in each covariate.

\begin{figure}
    \centering
\includegraphics[scale=.37]{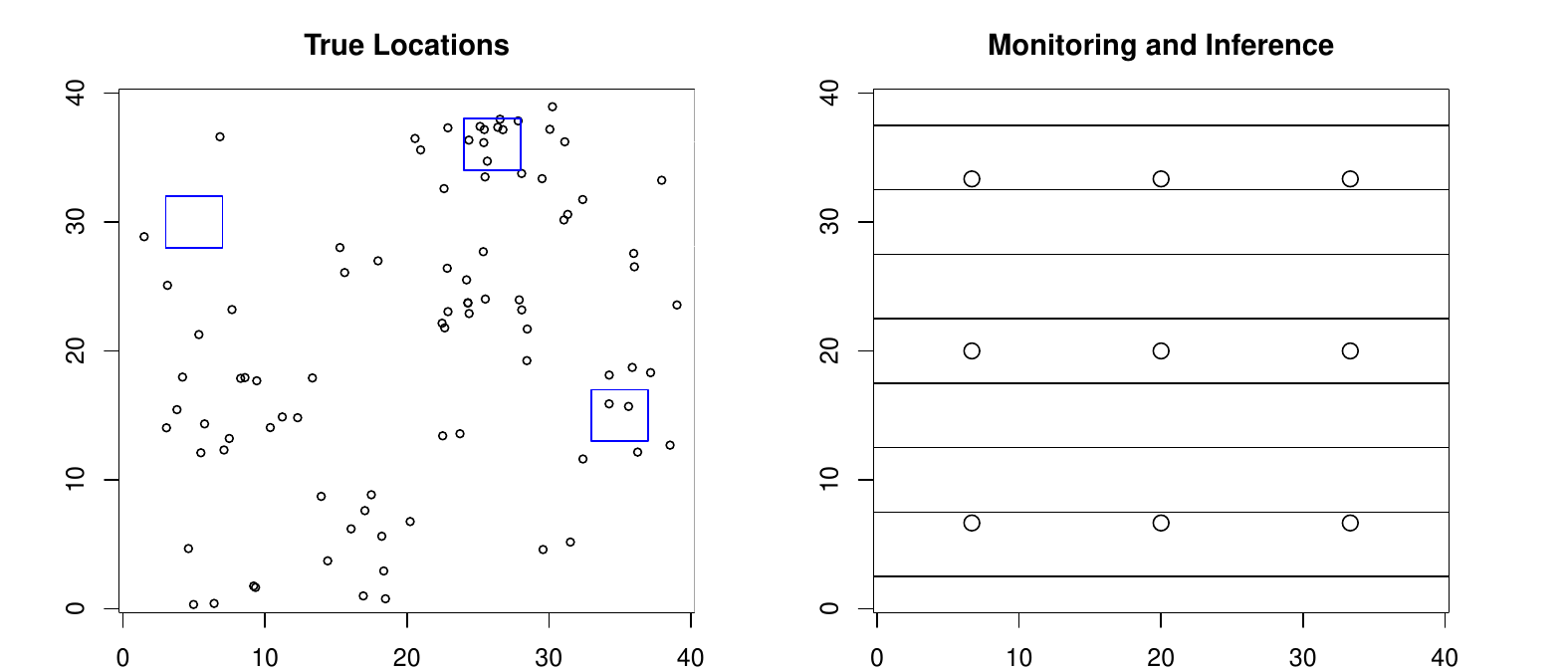}
    \caption{
    (left) True whale locations simulated using the true intensity surface. Also shown are three small regions for which inference will be conducted. (right) Monitoring design shows 8 aerial transects (horizontal lines) as well as 9 hydrophone locations. }
    \label{fig:True}
\end{figure}

The true values for the coefficient parameters are $\boldsymbol{\beta} = (\beta_0, \dots, \beta_3) = (-3.8, 0.3, 0.6, 0.9)$.
The true resulting intensity surface associated with this specification using a realization of each of the GPs is shown in Figure \ref{fig:sec2int} (top left).
Given this intensity, we then simulate a realization of the point pattern \citep{gelfand2018bayesian}. The simulated point pattern of true whale locations is shown in Figure \ref{fig:True} (left). This realization resulted in a total of 89 whales which we take as the unobserved truth.

\subsection{Distance Sampling}

Plane transect routes are fixed and known \citep{mayoDistributionDemographyBehavior2018}. Distance-based detection probability functions are also assumed known \citep{ganleyWhatWeSee2019} and are associated with a transect route, denoted by $p_{dist,\ell}(\bs)$ for route $\ell$.
We first introduce the notion of a whale being available to be seen in that it is at the surface at the time of observation.
While this probability is likely spatially and temporally varying based on behavioral patterns of the whale \citep{ganleyWhatWeSee2019}, we treat it as constant given insufficient data to inform this probability.
In practice, the probability of a whale being on the surface, denoted $\pi$, could be fixed based on scientific knowledge or treated as a random variable in the model and informed by auxiliary data.
The values of $\pi$ employed here are derived from tagging data in Cape Cod Bay (S. Parks, \textit{pers comm}).
As a result, $p_{dist,\ell}(\mathbf{s}) = \pi f(d(\mathbf{s},\ell))$
where $f$ is a distance based detection probability and $d(\mathbf{s},\ell)$ is the distance between location $\mathbf{s}$ and transect $\ell$.
The detection function for distance sampling we adopt follows from \cite{ganleyWhatWeSee2019}. Whales on the surface are detected with probability 1 if they are within 0.75km of the aerial transect. At distances further than 0.75km, the detection probability decays exponentially. That is,
\begin{equation}
f(d(\mathbf{s},\ell)) =
\begin{cases}
  1  & d(\mathbf{s},\ell) \leq 0.75km \\
  \exp(-(d(\mathbf{s},\ell)-0.75)^2) & d(\mathbf{s},\ell) > 0.75km
  \end{cases}
\label{eq:ds}
\end{equation}


In the simulation, we assume $L=8$ east-west aerial transects spaced 5km apart.
These aerial transects are shown spatially in Figure \ref{fig:True} (right).
Applying the detection process to the true whale locations generates an \emph{observed} distance sampling dataset for each transect $\ell$.
For each whale, we independently and randomly generate a binary variable to denote detected (1) or undetected (0) by each transect.  The binary variable is generated using a Bernoulli distribution where we assume a surface probability of $\pi = 0.4$ and detection function evaluated using the distance between the location of the whale and the nearest point along the transect.
In this simulation, 30 of the 89 whales were detected by the aerial transects.

\subsection{Passive acoustic monitoring data}

Hydrophones detect calls independently with probability based on the source level of the call, the ambient noise, and the distance between the hydrophone and where the call was made (i.e., the whale's location).
Let $p_k(\mathbf{s})$ to be the probability of detection of a call made at location $\mathbf{s}$ by hydrophone $k$ located at $\mathbf{h}_k$.
For illustration, we assume a fixed ambient noise level of 104 dB re:1$\mu$Pa with uniformly distributed upcall source level (SL) having lower and upper bounds of 141 and 197 dBs \citep{hatchQuantifyingLossAcoustic2012}.
Calls must be received on the hydrophone at 26 dB above ambient to be detected \citep{palmerAccountingLombardEffect2022a}.
The transmission loss coefficient is 14.5 and the transmission loss function is
$14.5 \text{log}_{10} (d)$
where distance $d$ is in meters.
Altogether, we adopt

\begin{equation}
\begin{split}
p_{PAM,k}(\mathbf{s}) &= P(SL - 14.5 \text{log}_{10} (d(\mathbf{s}, \mathbf{h}_k)) > 104  + 26)\\
 & =1-P(SL - 14.5 \text{log}_{10} (d(\mathbf{s}, \mathbf{h}_k)) \leq 104 + 26)\\
&=\begin{cases}
0 & 26+ 104+ 14.5\text{log}_{10}(d(\mathbf{s}, \mathbf{h}_k)) >197\\
1 & 26+104 + 14.5\text{log}_{10}(d(\mathbf{s}, \mathbf{h}_k))  <141\\
1- \left(\frac{26+104+ 14.5\text{log}_{10}(d(\mathbf{s}, \mathbf{h}_k))   - 141}{197 - 141} \right) & else \\
\end{cases}
\end{split}
\label{eq:am}
\end{equation}

The degraded intensity function for the acoustic data is written as a function of the intensity $\lambda(\mathbf{s})$ and the detection function $p_{PAM,k}(\boldsymbol{s})$.
Specifically, for hydrophone $k$, the expected number of calls is defined
$$\lambda_{PAM,k} = c\int_{D} \lambda(\mathbf{s})p_{PAM,k}(\boldsymbol{s}) d\mathbf{s}$$
where $c$ is the average number of calls made per whale. The number of calls detected by hydrophone $k$ is modeled as $Y_k \sim Poisson(\lambda_{PAM,k})$.

To simulate the passive acoustic monitoring data, we first simulate the number of calls that each whale makes during the specified time window.
In practice, $c$ could be fixed based on scientific knowledge or treated as a random variable in the model and informed by auxiliary data.
We simulate the numbers of call numbers independently from a Poisson distribution with intensity equal to the average number of calls parameter set at $c=6$, which is motivated by ancillary data \citep{parksSoundProductionBehavior2011}.
The number of calls in the simulation ranged from 0 to 12 per whale.

For acoustic detection, we assume a 3 $\times$ 3 array of hydrophones equally spaced across the region.
These hydrophone locations are shown spatially in Figure \ref{fig:True} (right).
The observed acoustic recording data are simulated based on the detection function and whale locations.
That is, for each hydrophone and for every call originated from every whale's location, we independently simulate a binary random variable which identifies whether or not the hydrophone detects the call based on its associated detection probability. Summing over these binary random variables for each hydrophone results in the total number of calls detected at each hydrophone location.
In the simulation, the number of recordings per hydrophone ranged from 39 to 71 with a total of 478 recorded calls.

\subsection{The probability of detection surfaces}

We present the probability of detection surfaces associated with each source as well as with the fusion.  These probabilities are not obtained from model fitting, but rather are explicit functions of the detection functions associated with the aerial distance sampling and PAM.
For the aerial data, the probability of observing a whale at location $\mathbf{s}$ for any $\mathbf{s} \in \mathcal{D}$, given that a whale is there and at the surface is computed $p_{dist}(\bs) = 1- \prod_{\ell}(1-p_{dist,\ell}(\bs))$. That is, it is the probability of detection by at least one trajectory route.
For the PAM data, the probability of a call being made by a whale at location $\bs$ and being heard by at least one hydrophone is equal to $p_{PAM}(\bs) = 1- \prod_{k}(1-p_{k}(\bs))$. Then, the joint probability of detection at $\bs$ from either source is $1-(1-p_{dist}(\bs))(1-p_{PAM}(\bs))$, which is greater than or equal to both $p_{dist}(\bs)$ and $p_{PAM}(\bs)$.  Following Section \ref{sec21}, we therefore expect the benefit from the data fusion to arise as a reduction in uncertainty in our inference on abundance.

\begin{figure}
    \centering
\includegraphics[scale=.55]{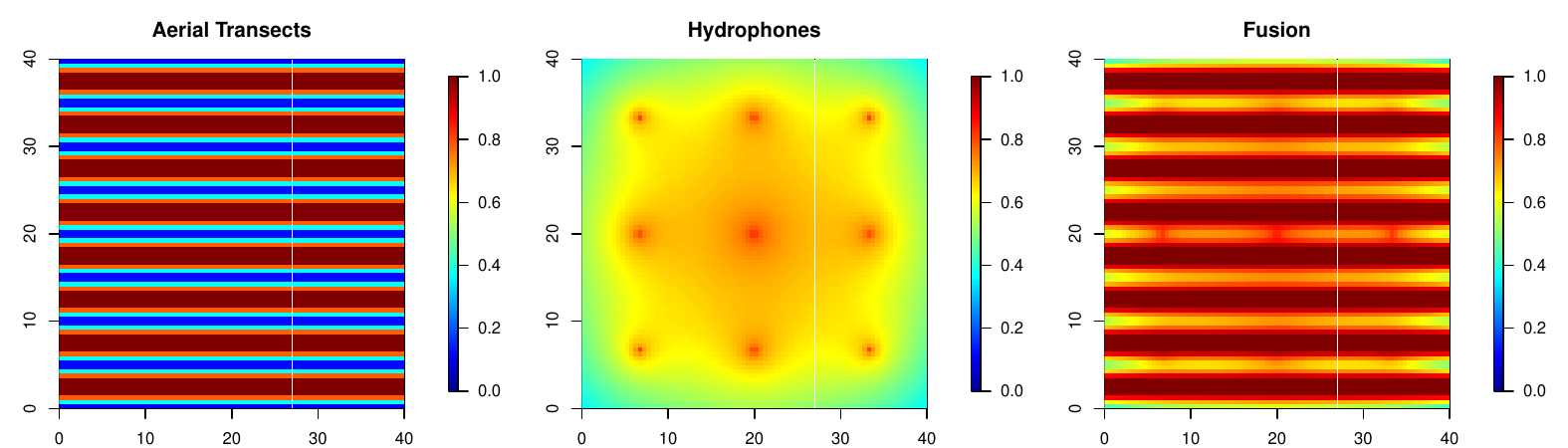}
    \caption{Detection probabilities based on moderate aerial transect sampling intensity (left), moderate hydrophone sampling intensity (center), or both (right) given that the whale is on the surface and calling.}
    \label{fig:detectionA}
\end{figure}

Figure \ref{fig:detectionA} shows the probability of detection surfaces for the aerial data only (left), the PAM data only (middle), and the data fusion (right) based on 8 aerial transects and 9 hydrophones.  
Under the aerial distance sampling design, the detection probability surfaces are fairly good, with the fusion surfaces offering improvements between transects. Importantly, both sources provide gains to detection since aerial detection is only possible when a whale is on the surface and PAM detection is only possible when a whale is calling.
Figure \ref{fig:detection} depicts the difference between the detection probabilities under the fusion and each data source separately.

\subsection{Simulation results}

Given the likelihood functions in (\ref{eq:liksource1}) and (\ref{eq:liksource2}), the two single data source models as well as the data fusion model were fitted to the simulated data.
Each of the models was fitted within the Bayesian framework.
Prior distributions were assigned to the model parameters.
We assumed an intercept-only model, where $\beta_0$ was assigned an independent Normal prior with mean 0 and variance 1000$^2$. The spatial variance parameter, $\sigma^2$ was assigned a conjugate Inverse Gamma prior with shape and scale of 2. In the simulation, the range parameter, $\phi$, of the exponential covariance function was fixed at 3.

In attempting to estimate the magnitude of $\lambda(\bs)$, only the surface probability parameter, $\pi$ or the call rate parameter, $c$, can be identified from the two data sources. In the simulation, we fix these values to their true values, $\pi = 0.40$ and $c = 6$. An incorrect specification of these values results in a uniform scaling of the resulting intensity surface, $\lambda(\bs)$ but the spatial variation in the intensity surface is retained. In the application, we are able to learn these two parameters given ancillary data and additional modeling. See Section \ref{sec4} for further details.




Markov chain Monte Carlo and a Metropolis-within-Gibbs sampling algorithm were used to obtain posterior inference. For each model, we obtained 20000 posterior samples. The first 5000 were discarded as burn-in and the remaining 15000 were used for posterior inference. Trace plots were examined and no issues of convergence were detected.

\begin{figure}
    \centering
\includegraphics[scale=.55]{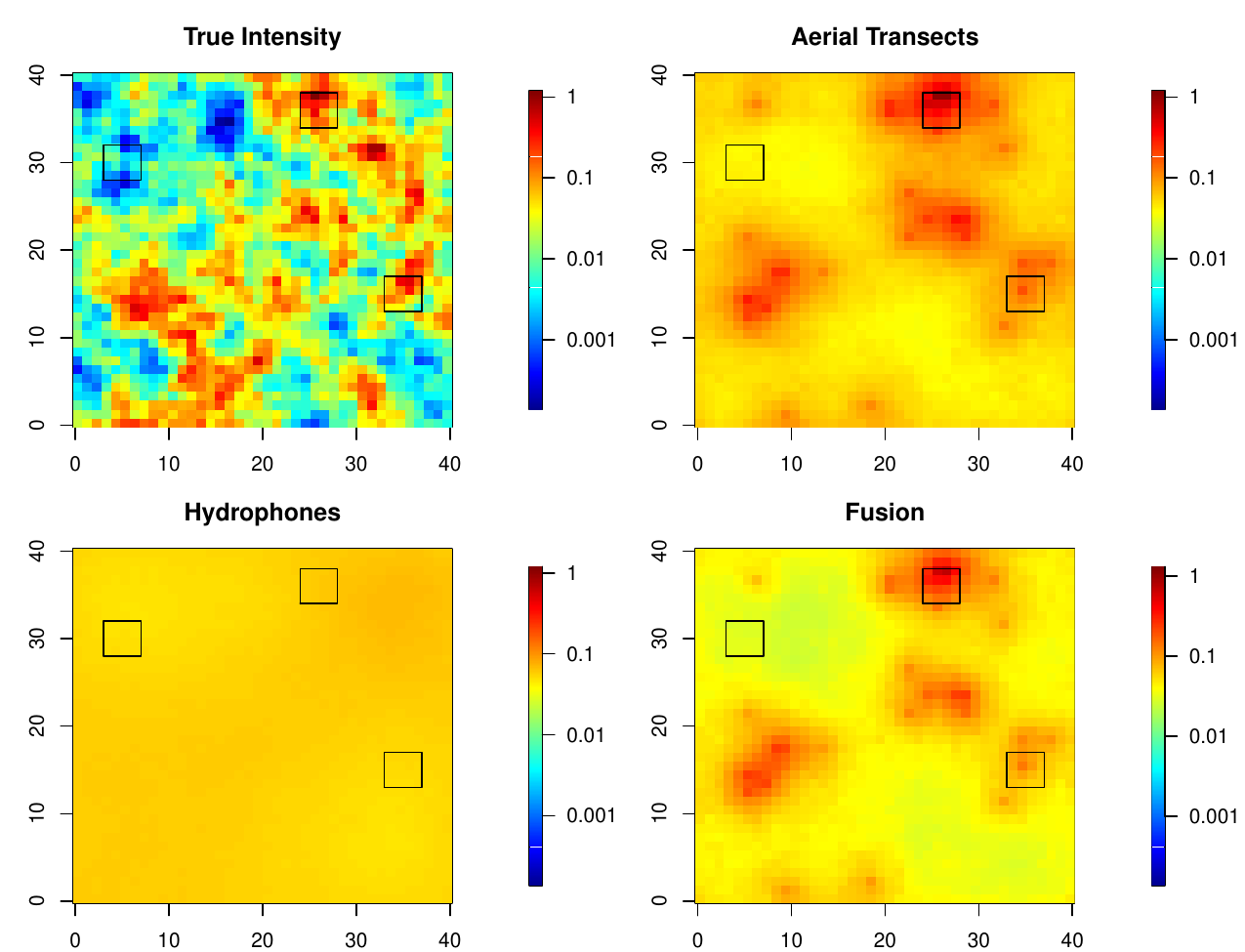}
    \caption{True intensity surface (top left) as well as the posterior mean estimated intensity surface for the aerial transect data only model (top right), hydrophone data only model (bottom left), and data fusion model (bottom right). }
    \label{fig:sec2int}
\end{figure}

Posterior mean estimates of the intensity surface are shown in Figure \ref{fig:sec2int} for each of the models. Overall, the estimated intensity surfaces depict similar trends across the region showing higher intensity regions in the lower left and mid to upper right as well as a lower intensity region in the upper left and lower right. In this simulation, the model fitted with only PAM data resulted in the smoothest posterior mean intensity surface. Comparatively, the estimates from the model using only the distance-based aerial sampling data have sharper contrasts between the higher and lower intensity regions. The data fusion model using both sources results in the most precise estimated intensity surface.

\begin{figure}
    \centering
\includegraphics[scale=.45]{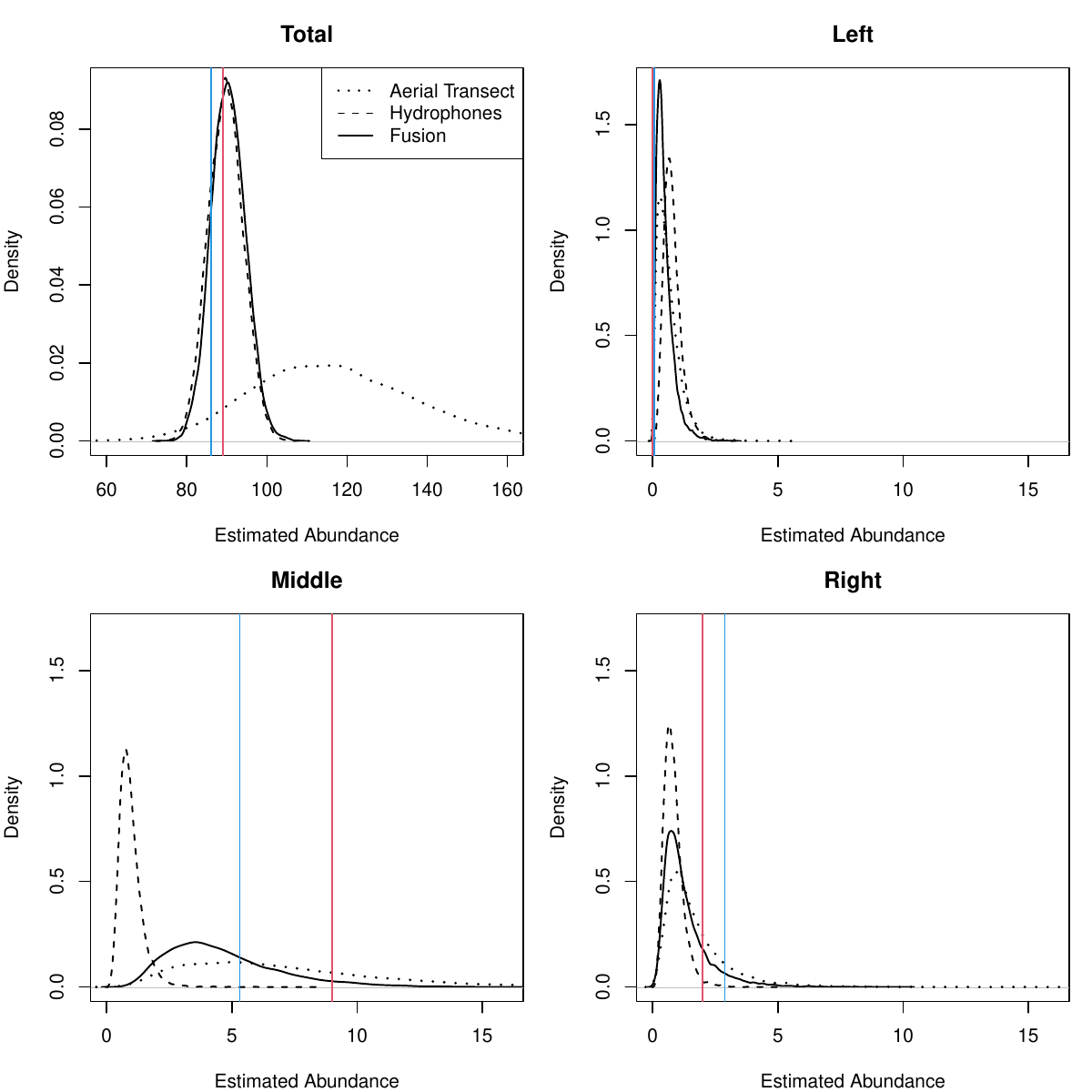}
    \caption{Posterior density estimates of total abundance (top left), as well as the estimated abundance for each subregion, left (topright), middle (bottom left), and right (bottom right). The red vertical lines represent true simulated abundance and the blue vertical lines represent expected abundance, each computed for the entire region and three subregions.}
\label{fig:densitycurves}
\end{figure}

For each model, we obtained posterior estimates of total abundance, computed as $\int_{\mathcal{D}} \hat{\lambda}(\mathbf{s})d\mathbf{s}$. Posterior distributions are shown in Figure \ref{fig:densitycurves}. The red vertical line represents true total simulated abundance (89 whales) and the blue vertical line depicts the expected total abundance (86.10 whales), computed as $\int_{\mathcal{D}} \lambda_{true}(\mathbf{s})d\mathbf{s}$.
Posterior mean and standard deviations of total abundance are given in Table \ref{table:sec2est}, where each model recovers the true value of 89 whales within the 95\% credible interval. The fusion model is the most accurate in terms of both the posterior mean estimate of total abundance and the smallest posterior standard deviation.

\begin{table}[ht]
\caption{Posterior mean (standard deviation) estimates of total abundance for each of the three models. The true abundance is 89 and the expected abundance is 86.10. Also included are the RMSE, log posterior density, and RPS. \label{table:sec2est}}
\centering
\begin{tabular}{lcccc}
  \hline
    &&&Log posterior\\
  & Abundance &RMSE&density &RPS\\
  \hline
Aerial Transects     & 117.85 (21.33) & 1.90& -348.97 (12.56) & 18.08 \\
Hydrophones         & 89.67 (4.53) & 1.89& -374.14 (19.67) & 1.08\\
Fusion              & 89.76 (4.40) & 1.65 & -348.29 (14.59) &1.06\\
   \hline
\end{tabular}
\end{table}

Using the discretized true and estimated intensity surfaces, we computed the root mean square error (RMSE) of the log intensity for each model. In addition, the log posterior density and rank probability score was also computed for each model. These values are also reported in Table \ref{table:sec2est}. Across the entire domain, the fusion model outperformed both single source models based on each of the three measures. The aerial transect data only model was similar to the fusion model based on the log posterior density measure while the hydrophone data only model was similar to the fusion model based on rank probability score (RPS).


For the left, middle, and right subregions, we also estimated total abundance for each of the three models (Figure \ref{fig:densitycurves} and Table \ref{table:sec2estsub}). Each of the models performed similarly in estimating abundance for the left region where true abundance was 0. In the middle and right regions, the fusion and aerial transect data only model better captured the true values of 9 and 2, respectively.
Note that the expected abundance in the three regions was 0.07 (left), 5.31 (middle), and 2.88 (right). Given the smoothness of the estimated intensity surface using only the hydrophone data, this model performed the worst across all three subregions despite being able to accurately estimate total abundance for the region.
The data fusion model outperformed both of the single source models for each of the subregions based on the RMSE (Table \ref{table:sec2estsub}).

\begin{table}[ht]
\caption{Posterior mean (standard deviation) estimates of abundance as well as RMSE for the three subregions shown in Figure \ref{fig:sec2int} for each of the three models. True abundance for the three regions was 0 (left), 9 (middle), and 2 (right), and the expected abundance in the three regions was 0.07 (left), 5.31 (middle), and 2.88 (right).  \label{table:sec2estsub}}
\centering
\begin{tabular}{lccc}
  \hline
 & Left & Middle & Right\\
 \hline
\multirow{2}{*}{Aerial Transects}  & 0.63 (0.55) & 8.02 (4.01) & 1.98 (1.58) \\
& 2.91 	& 1.21 			& 1.29  \\
\multirow{2}{*}{Hydrophones} & 0.74 (0.40) & 0.97 (0.63) & 0.80 (0.52)\\
&3.08 & 1.60 & 1.41 \\
\multirow{2}{*}{Fusion}  & 0.36 (0.32) & 6.13 (2.74) & 1.49 (1.13)\\
& 2.40 & 0.97 & 1.19\\
   \hline
\end{tabular}
\end{table}


\subsection{Simulation study for novel data fusion}
We further investigate the performance of our proposed data fusion model through a fairly broad simulation. We consider multiple levels of (i) intensities of sampling designs specified by the number of hydrophones and aerial transects, (ii) surface probabilities, $\pi$, (iii) the average number of calls, $c$, and (iv) total abundance. For each combination of the above, we fit the data fusion model as well as the comparable single data source models. We compare the models based on their ability to recover the true intensity surface of abundance and estimate total abundance for the region.

Full details on these simulation results are provided in the supplementary material, including intensity surface estimates, estimates of total abundance, and RMSEs.
All models capture the true value of abundance based on the 95\% credible intervals.
For fixed sampling intensities, surface probabilities, and call rates, RMSE estimates scale with abundance.
Uncertainty in the estimated surfaces and in total abundance decrease as whales become more detectable -- whales on the surface with higher probabilities and having higher call rates.
In general, model performance increases with increases in the sampling intensity of either or both data sources.
Higher PAM sampling or higher aerial transect sampling results in more refined estimates of the intensity surface, estimates of abundance, and lower RMSEs.
For a fixed aerial transect sampling intensity, the difference between the low, moderate, and high PAM sampling intensity is modest compared to the the difference between low, moderate, and high aerial transect sampling for a fixed PAM sampling intensity (Figure \ref{fig:Est}). 
In terms of estimating total abundance, the PAM data appears to be slightly more valuable than the aerial transect data data.
See the supplementary material for further results and discussion.


A takeaway from the foregoing simulation is the possibility to use simulation to develop sampling design.  For example, it could be used to investigate where and when to do aerial distance sampling as well as how many PAMs to use and where to place them. Together, this would inform about the trade-off between the two collection modes in learning about true intensity.

\section{Abundance estimation of the NARW in CCB}
\label{sec4}

We apply our fusion model to data collected on North Atlantic right whales in Cape Cod Bay on April 9 and 10, 2009. Distance-based aerial sampling surveys were flown on both days and the PAM data collected during the time window of each flight were extracted (See Yack et al., \textit{In prep}, for details on the passive acoustic data). On April 9, seven transects in the southern region of CCB were flown between 5:18pm and 8:45pm UTC. Eight whales were detected through the distance sampling survey and their locations are depicted as {\color{blue} $\times$} in Figure \ref{fig:CCBdata}. During this period, 1031 calls were recorded across the 10 PAMs. The sizes of the {\color{red} $\circ$} in Figure \ref{fig:CCBdata} are proportional to the number of calls heard on each monitor. On April 10, aerial surveys were flown between 1:55pm and 10:17pm UTC, during which 46 whales were seen across fifteen transects spanning the entirety of CCB. During this same time period, 486 calls were recorded on the hydrophones. These data are also depicted in Figure \ref{fig:CCBdata}.

Auxiliary data were obtained from high resolution tags deployed in April of 2009 and 2010 by S. Parks and D. Wiley (pers. comm). These data provided realizations of the number of calls per hour made by three whales during a sampling event. We modeled these data with a Gamma distribution distribution with mean equal to $c$ and variance 10. From the tag data, we also obtained dive data to inform about the surface probability parameter, $\pi$. Specifically, we extracted observations of the proportion of time on the surface based on the time spent within 2 meters of the surface for 15 whales. The average was 0.66 with minimum and maximum equal to 0.10 and 0.91, respectively. We modeled these data using a Beta distribution with $\mu=\pi$ and $\nu = 15$ based on the mean and sample size parameterization.

\begin{figure}
    \centering
\includegraphics[scale=.325]{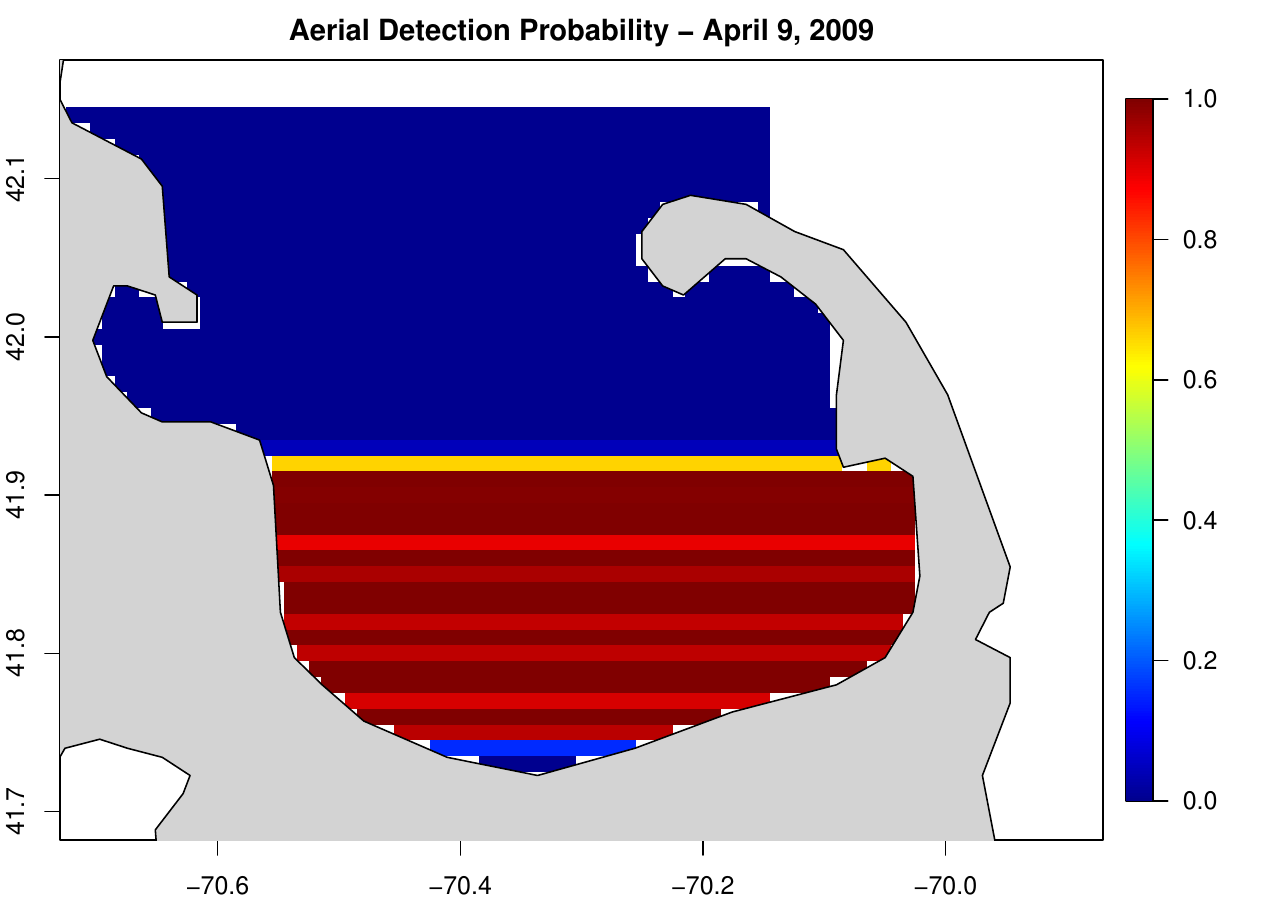}
\includegraphics[scale=.325]{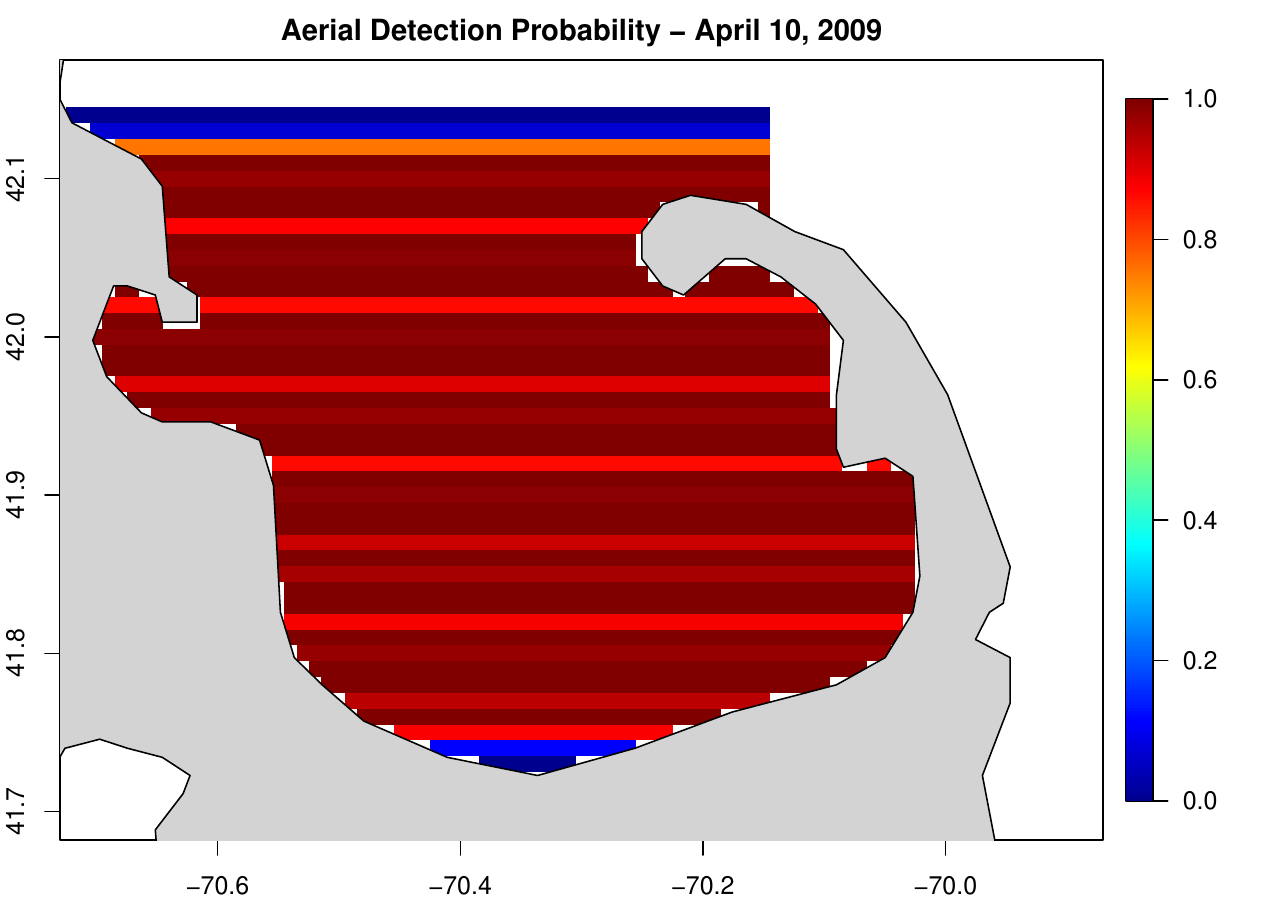}\\
\includegraphics[scale=.325]{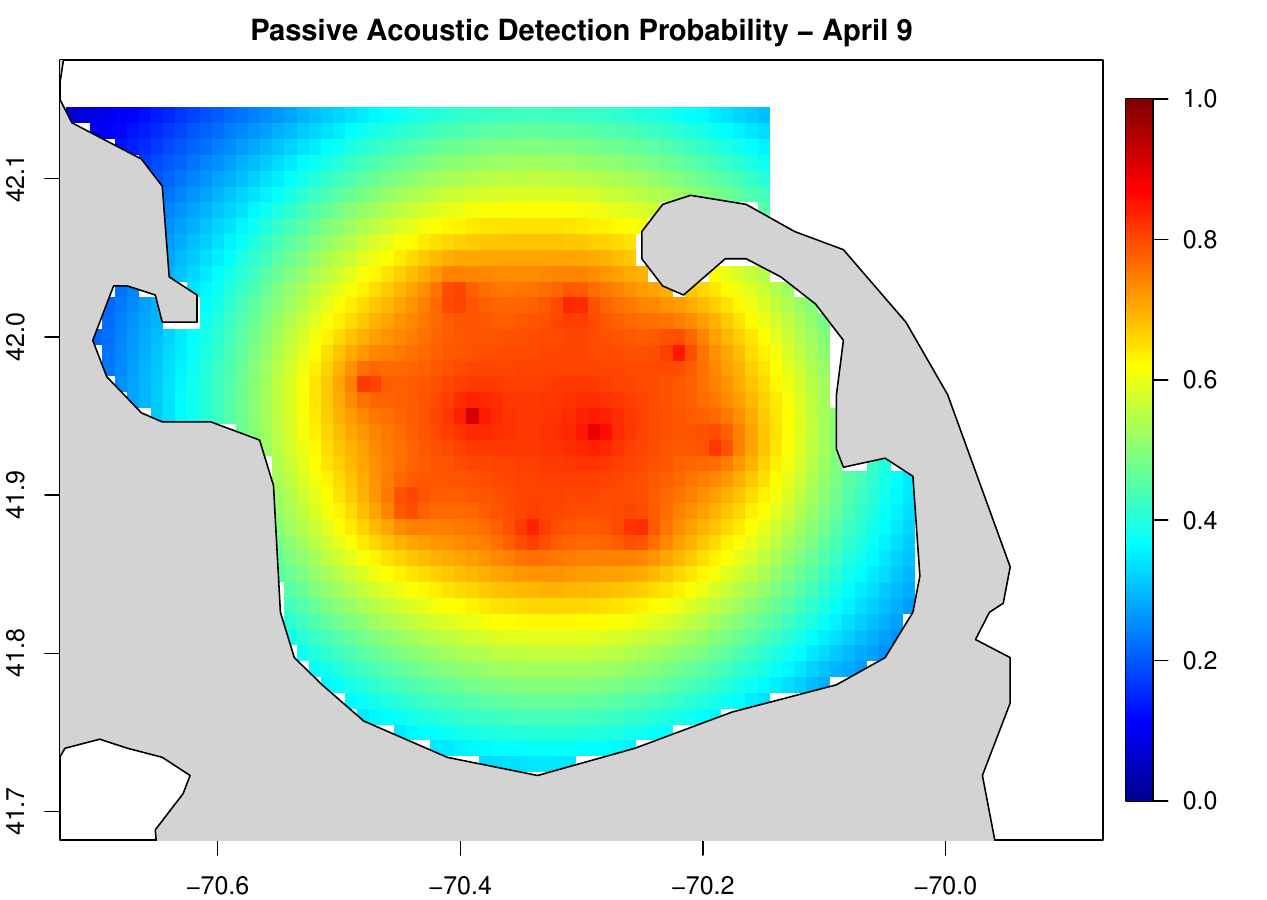}
\includegraphics[scale=.325]{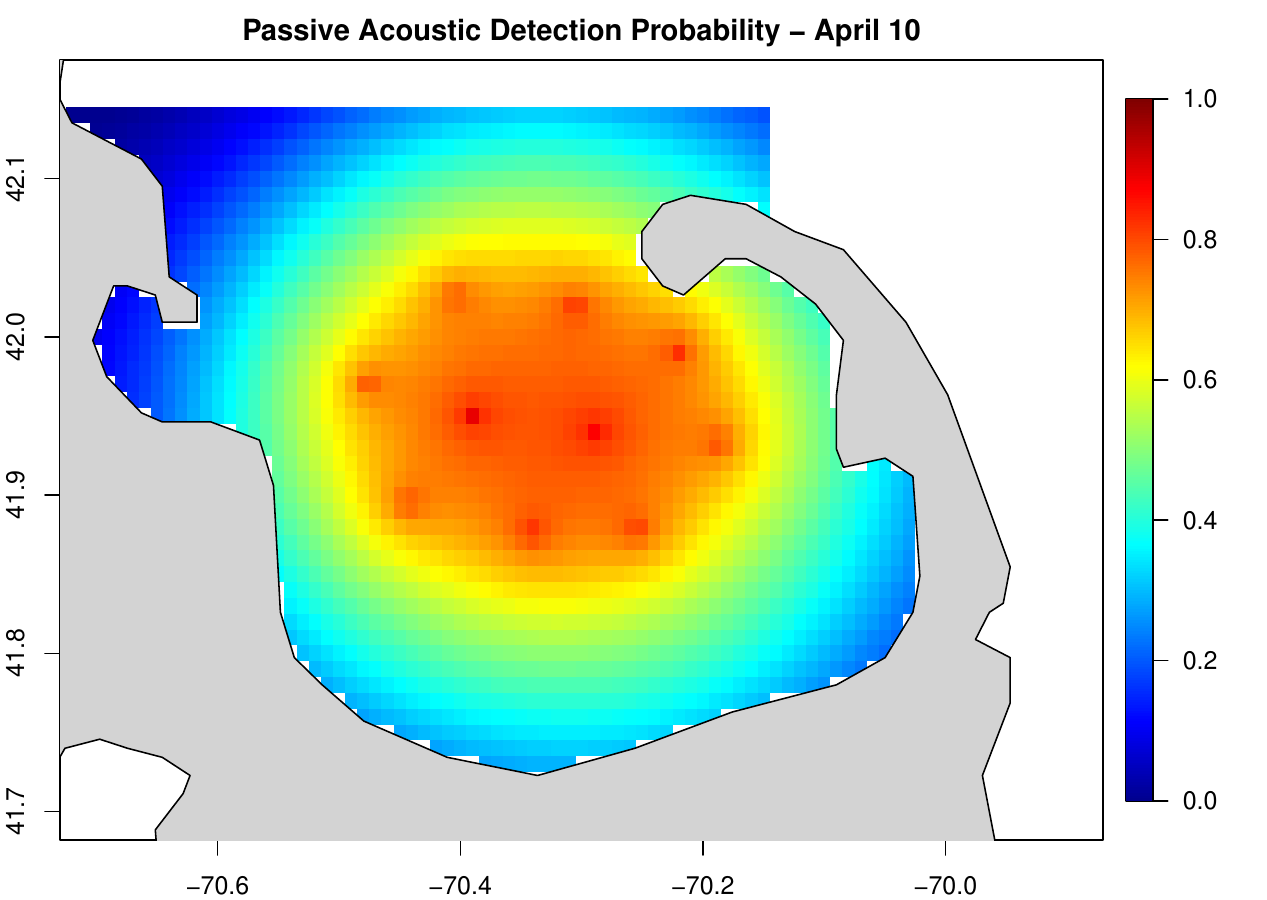}
    \caption{Detection probabilities for April 9 (left) and 10 (right) for aerial transects (top) and passive acoustic recorders (bottom). These probabilities are computed using the detection functions in (\ref{eq:ds}) and (\ref{eq:am}), conditional on a whale being at the surface at each location and making a call.}
    \label{fig:DetProb}
\end{figure}

We computed the detection probabilities across the region for April 9 and 10 for each source, which are conditional on a whale being at the surface at each location and making a call. These probability surfaces are shown in Figure \ref{fig:DetProb}. The aerial detection probability is high near the transects. On April 9, a day of incomplete coverage, non-zero detection probability is limited to the southern half of CCB.
In comparison, the hydrophone arrangement produces much more complete spatial coverage but with moderate detection probability.
These detection probabilities are computed using hydrophone- and day-specific ambient noise levels from CCB, which ranged between 102.9 and 108.1 dB re:1$\mu$Pa. The average ambient noise level recorded across the ten hydrophones was 105.2 dB on April 9 and 105.9 dB on April 10, resulting in slightly lower detection probabilities across the region on April 10.

\begin{figure}
    \centering
\includegraphics[scale=.325]{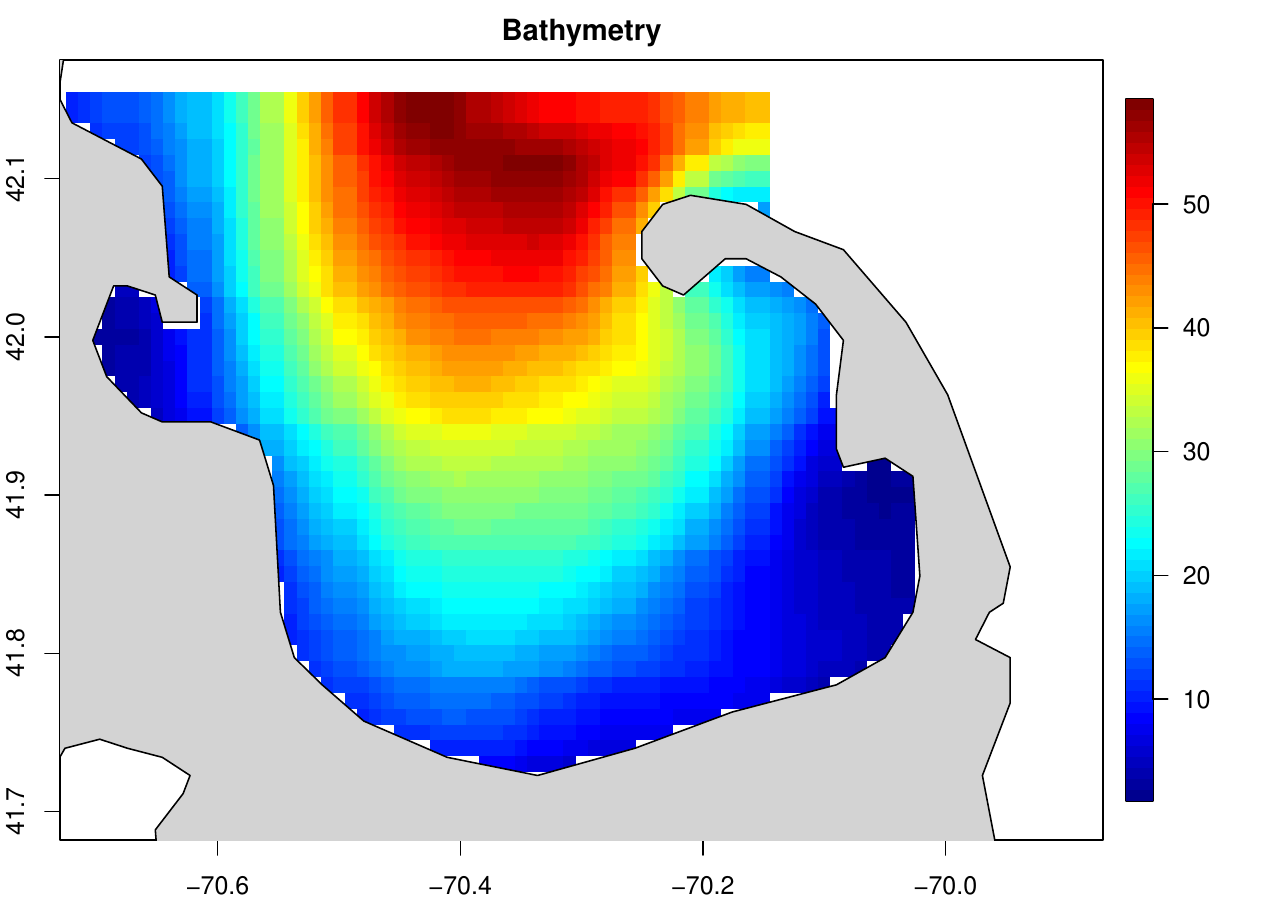}
    \caption{Bathymetry, reported in meters (m), throughout CCB.}
    \label{fig:Bath}
\end{figure}

For this analysis, we included bathymetry as a covariate in the LGCP model for $\lambda(\mathbf{s})$.
Bathymetry data were obtained from the SRTM Global Bathymetry and Topography dataset (\url{https://doi.org/10.5069/G92R3PT9}).
Figure \ref{fig:Bath} shows the bathymetry across CCB, depicting a strong north to south gradient in addition to shallower depths near the coastline.

We fitted the data fusion model to the two days of data separately.
For comparison, we also fitted an aerial transect data only model to the data collected on April 10 given the complete survey of CCB. We did not fit an aerial transect data only model to the data collected on April 9 since it would require vast extrapolation in the northern half of CCB.
Prior distributions were assigned to all model parameters.
The intercept, $\beta_0$, was assigned a non-informative Normal distributions with mean 0 and variance 100. The bathymetry coefficient, $\beta_1$, was assigned a Normal distributions with mean 0 and variance 1.
The spatial variance was assigned a Gamma distribution with shape and rate parameters equal to 2.
For computational efficiency, we fix the spatial decay parameter to an effective range equal to $\frac{1}{3}$ max distance, which is equal to $\approx$ 23km in CCB.
We investigated the sensitivity of this choice and found very little difference in terms of abundance estimates for the region.
For the surface probability parameter, $\pi$, we assign a Uniform(0,1) prior distribution and for the average call rate parameter, $c$, we assign a Uniform(0,100) prior distribution.

Model fitting was done through Markov chain Monte Carlo run for 100,000 iterations. The first 20,000 posterior samples were discarded as burn-in and the remaining samples were retained for inference. Trace plots were inspected for convergence and no issues were detected.

Posterior mean and standard deviation estimates for the fusion model parameters for each day are presented in Table \ref{tab:params}. On both April 9 and 10, we detect a negative relationship between the intensity of abundance and bathymetry, although neither is significant. The spatial variance on April 9 is much smaller than on April 10, which can be attributed to fewer sightings of whales yet many acoustic recordings. The average call rates per hour per whale are significantly different for the two days; we estimate 3.86 calls per hour per whale on the April 9 and 1.23 calls per hour per whale on April 10. We do not detect any difference between the surface probabilities on the two days.

\begin{figure}
    \centering
\includegraphics[scale=.325]{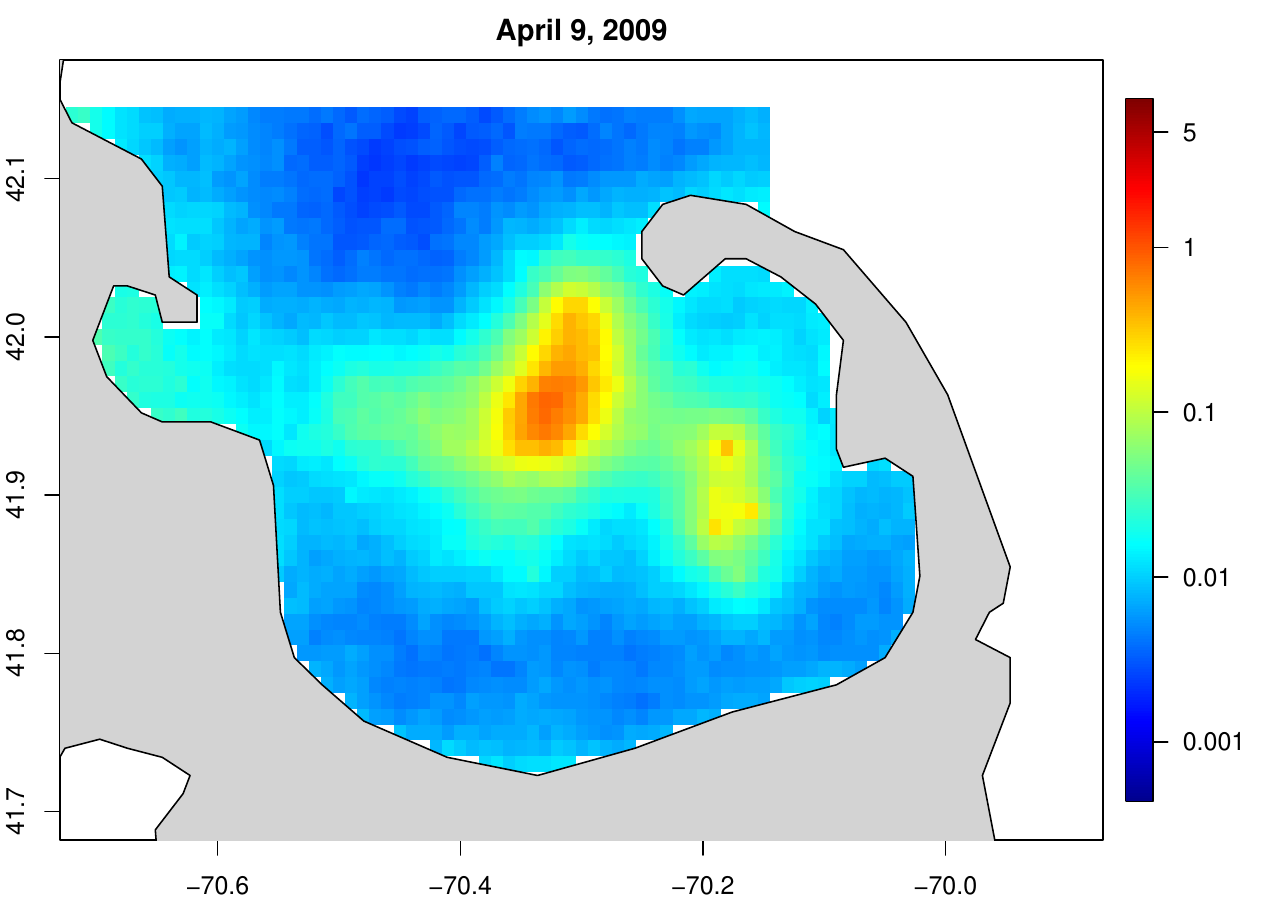}
\includegraphics[scale=.325]{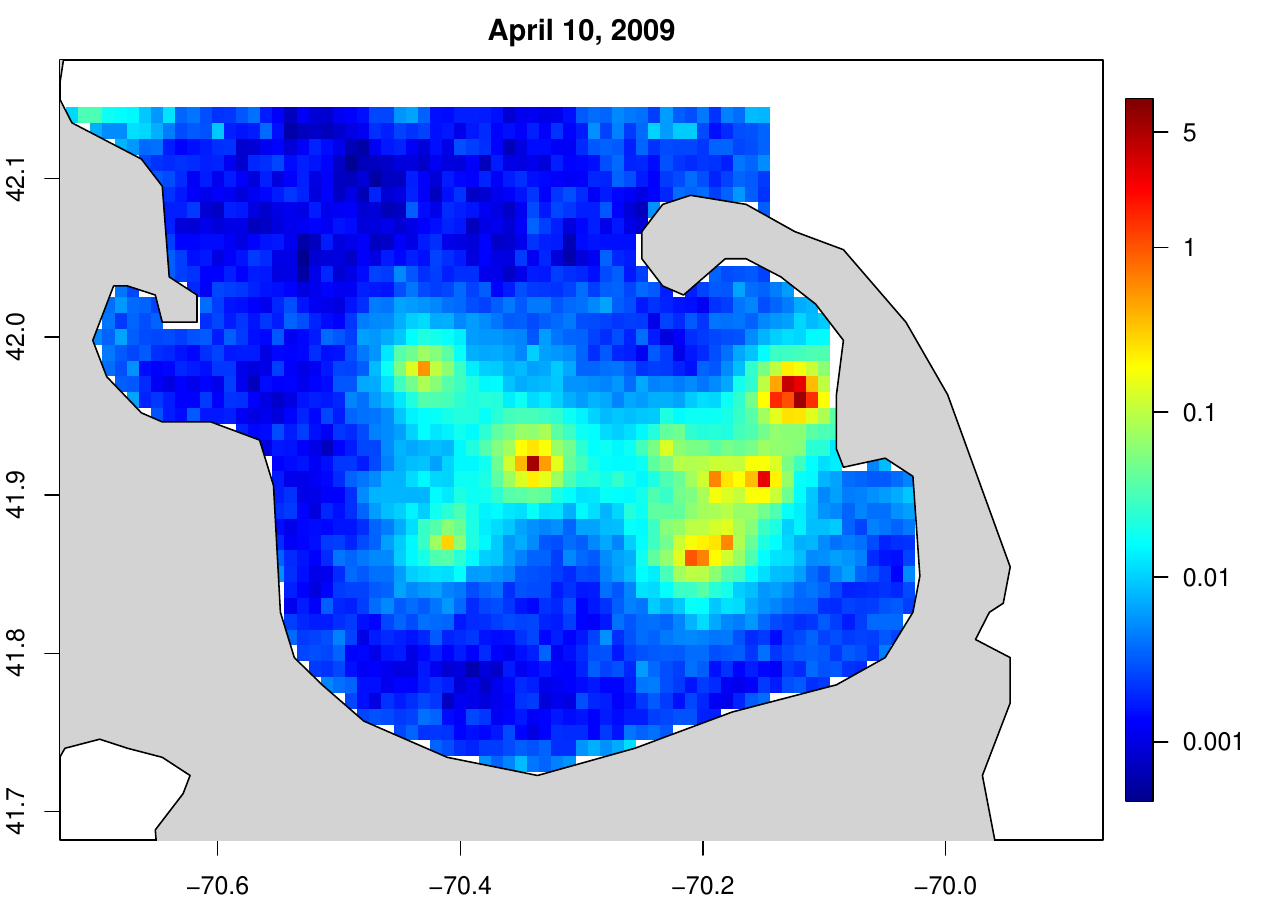}
    \caption{Posterior mean surface of the intensity of abundance for Cape Cod Bay on April 9, 2009 (left) and April 10, 2009 (right) under the proposed data fusion model.}
    \label{fig:EstApril}
\end{figure}

Figure \ref{fig:EstApril} shows the posterior mean estimate of the intensity surface for the two days. On April 9 we see two high density regions centrally located in CCB. Note that only 8 whales were seen during aerial distance sampling on this day, all of which were on the eastern side of the region. With over 1,000 calls detected on April 9, moderate intensity is estimated across much of central CCB, with a peak intensity near the center of the hydrophone array.  On April 10, high density clusters of whales are estimated across central and eastern CCB. Regions with very low estimated abundance are regions where distance sampling occurred and where no whales were detected. A notable difference between the intensity surfaces is the smoothness of estimates -- the surface is much smoother on April 9 than April 10. This can be attributed to the fact that on April 9, the whales were heard more than they were seen, and the acoustic data provide less information with respect to their precise locations. On April 10, they were seen more than heard, and given no false positive detections, the resulting intensity surface estimate is more precise for their locations.

The posterior distribution of total abundance across CCB is calculated for each day by integrating the intensity function over the spatial domain for each posterior sample of the parameters.
The posterior mean estimate of abundance on April 9 is 63.53 whereas the estimate on April 10 is 53.04. There were 7 behavioral observations made from the airplane on 9 April; 3 were of feeding whales (43\%). On April 10, there were 18 behavioral observations, 13 of which where of feeding whales (73\%). This corresponds to the anatomical evidence that feeding whales do not call. This points to the need to factor in both data collection modalities, as well as observed whale behavior to better understand how abundance may change over  longer periods.  

\begin{figure}
    \centering
\includegraphics[scale=.325]{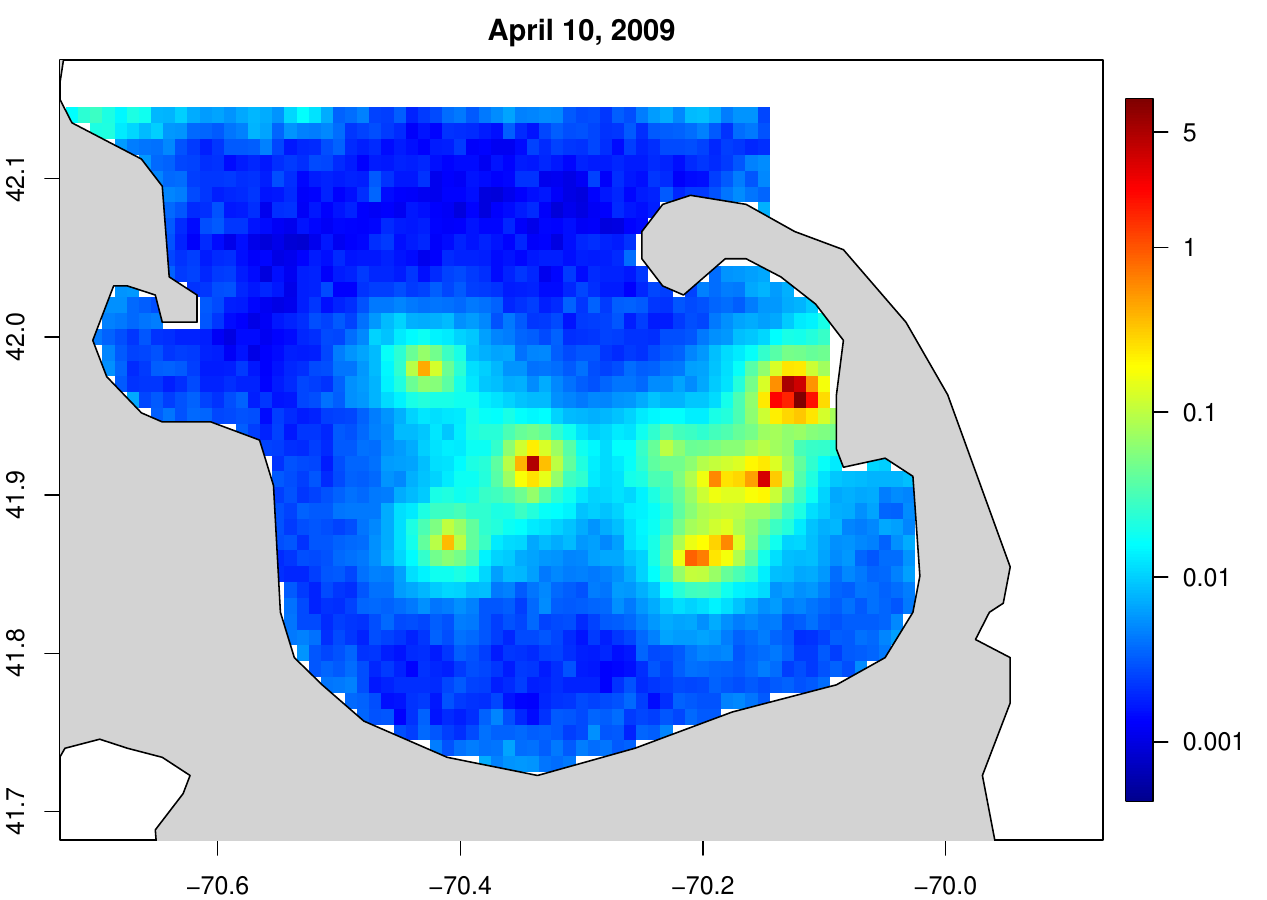}
    \caption{Posterior mean surface of the intensity of abundance for Cape Cod Bay on April 10, 2009 using only the aerial transect data.}
    \label{fig:EstApril10A}
\end{figure}

Lastly, we compare the results of the data fusion model to the model using only the aerial transect data for April 10. The parameter estimates obtained for April 10 using only the aerial transect data are similar to those from the fusion model (Table \ref{tab:params}) as is the posterior mean surface of the intensity of abundance (Figure \ref{fig:EstApril10A}).
The posterior mean estimate of total abundance is 61.88, which is slightly higher than that from the fusion model. This difference can partially be attributed to the intensity estimate under the fusion model being informed by the low volume of calls recorded on the PAMs. The uncertainty in the total abundance estimate as depicted by the posterior standard deviation is higher when using only the one data source. Whereas overall these results are similar to the fusion model, we are limited to the few days in which the aerial transects are flown and their coverage of CCB is complete.

\begin{table}[ht]
\centering
\caption{
Posterior mean (standard deviation) estimates for the model parameters for April 9 and 10, 2009 as well as estimates of abundance across CCB. \label{tab:params}}
\begin{tabular}{cccc}
  \hline
  &\multicolumn{2}{c}{Fusion}& Aerial transect only\\
 & April 9 & April 10 & April 10\\
  \hline
$\beta_0$ &  -1.18 (0.71) & -0.52 (0.85) & -0.59 (0.81) \\
$\beta_1$ & -0.38 (0.26) & -0.28 (0.36) & -0.33 (0.35) \\
$\sigma^2$ & 4.16 (0.86) & 11.06 (2.18) & 9.79 (1.14) \\
$c$ & 3.86 (0.87) & 1.23 (0.23) &    \\
$\pi$ &  0.65 (0.03) & 0.66 (0.03) & 0.65 (0.03) \\
Estimated & \multirow{2}{*}{63.53 (14.97)} &\multirow{2}{*}{53.04 (8.93)}&\multirow{2}{*}{61.88 (9.74)}\\
Abundance  &  &  \\
   \hline
\end{tabular}
\end{table}

\section{Summary and future work}
\label{sec5}

We have considered a novel data fusion problem of learning about marine mammal distribution and abundance. The two data sources uniquely provide partial information on abundance -- the aerial distance sampling data produces a partially observed point pattern while the PAMs collect a partial sample of whale calls associated with a network of hydrophones. 
Though the true intensity will never be known, through simulation and with real data from Cape Cod Bay in Massachusetts in the U.S., we have been able to demonstrate the benefit of the data fusion in terms of improved accuracy and precision.

Results presented here focus on the static spatial setting.  In practice, real data will exhibit misalignment in space and time. Assuming a common domain for the sources, we will need to model process dynamics in order to extend this modeling framework to the spatio-temporal context. Introducing $\lambda_{t}(\bs)$ requires the choice of temporal scale for which the dynamics in the intensity are suitable. Future considerations include which environmental features will drive the dynamics of the intensity. For example, given that NARW are known to be in CCB to feed on copepods \citep{hudakNorthAtlanticRight2023}, incorporating information on their food sources could aid in these modeling efforts. Additionally, there is a behavioral component to acoustic calling \citep{franklinUsingSonobuoysVisual2022}, so furthering the specification of the call rate process in the dynamic model will be necessary. The ability to make annual abundance estimates is a motivation for this ongoing research.

With continuing sampling technology advances, another area of further investigation considers fusion with additional data sources \citep{bucklandWildlifePopulationAssessment2023}.  For instance in CCB, we have collected opportunistic sightings of NARW co-incident with zooplankton surveys \citep{hudakNorthAtlanticRight2023}. Future methodology will be required to enable these data sources to inform on marine mammal distribution and abundance.
Finally, given the complexity of studying a species that spends much of its time unavailable for study, we are interested in learning about the relative balance of design and deployment of different data collection programs to better discern spatially explicit abundance.

\begin{funding}
   Aerial survey data for this study were collected under NOAA Scientific Permit No. 633-1763. CCS acknowledges funding from NOAA Fisheries, Massachusetts DMF, Massachusetts Environmental Trust, and private donors. The DTAG data were collected under NOAA NMFS Scientific Permit nos. 655-1652-01, 775-1875 and 633-1763 and were approved by the Penn State Institutional Animal Care and Use Committee. Funding for DTAG data collection work was provided by: US Office of Naval Research grants N00014-08-1-0630 and N00014-09-1-0066. This work was supported in part by funding from NOAA Fisheries under award NA20NMF0080246, SERDP award RC20-1097, and US Office of Naval Research Award N000141712817. 
\end{funding}

\begin{supplement}
\stitle{Simulation study}
\sdescription{Additional details and results pertaining to the simulation study.}
\end{supplement}

\bibliographystyle{apalike}
\bibliography{fusion}

\FloatBarrier


\newpage 
\title{Supplementary material}

\renewcommand*{\thefigure}{S\arabic{figure}}
\let\figurename\relax
\setcounter{figure}{0}

\renewcommand*{\thetable}{S\arabic{table}}
\setcounter{table}{0}

\begin{figure}
    \centering
\includegraphics[scale=.55]{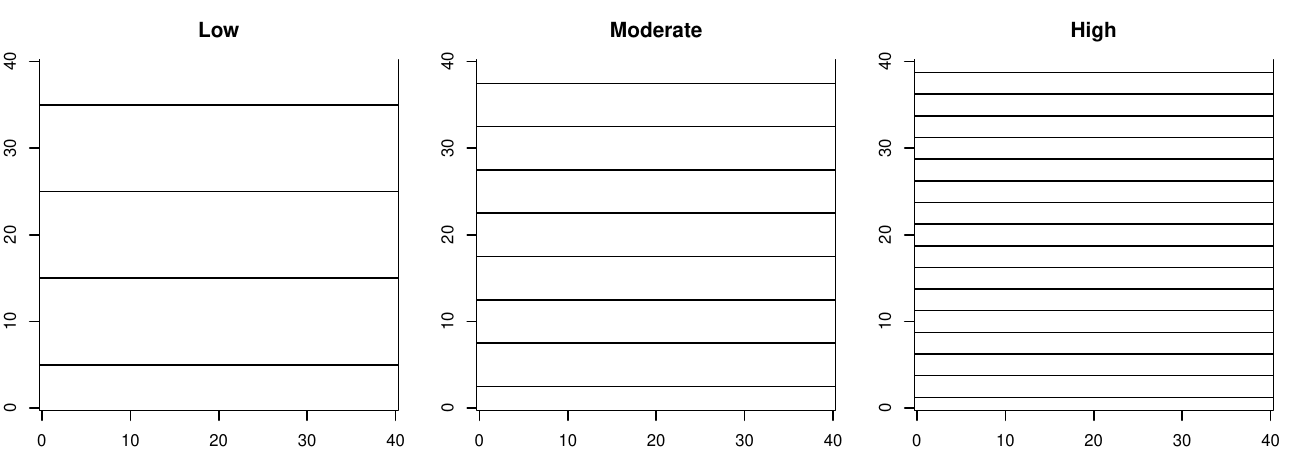}
\includegraphics[scale=.55]{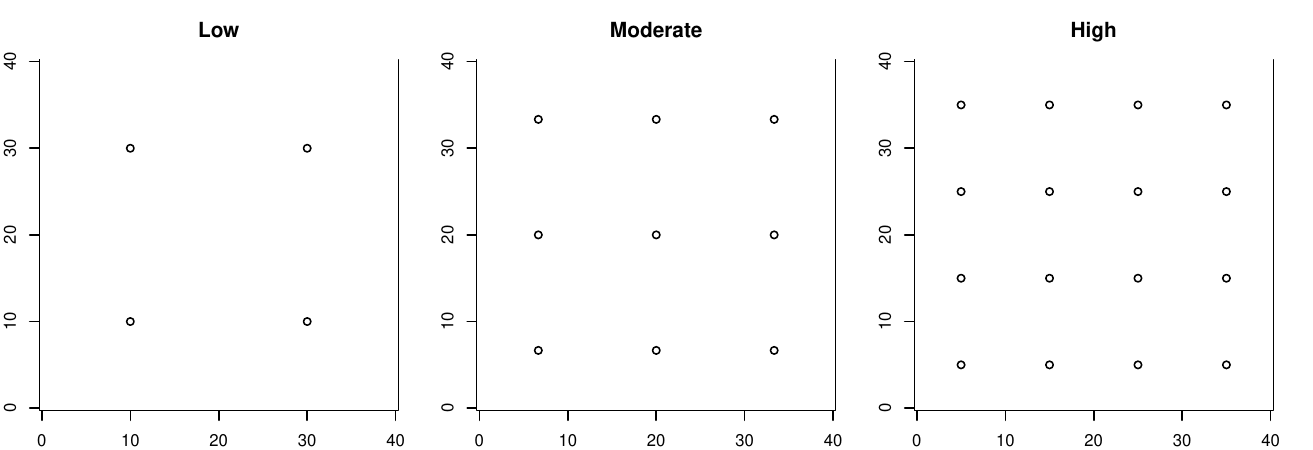}
    \caption{(top) Aerial transects flown under low, moderate, and high sampling intensity. (bottom) Hydrophone locations under low, moderate, and high sampling intensity.}
    \label{fig:aerialhydro}
\end{figure}

We continue our investigation of the performance of our proposed data fusion model through a broad simulation.
First we consider three different degrees of sampling intensities for both data sources, nothing that \emph{moderate} sampling intensities were used in Section \ref{sec3}.
For distance sampling, we consider low, moderate, and high sampling intensities defined by 4, 8, and 16 equally spaced aerial transects across the domain $\mathcal{D} = [0,40] \times [0,40]$. These flight paths are shown in Figure \ref{fig:aerialhydro}.
Similarly, for acoustic detection, we consider three space-filling designs based on a $2 \times 2$, $3 \times 3$, and $4 \times 4$ grid, which are denoted low, moderate, and high sampling intensity, respectively. The locations for each of the designs are also shown in Figure \ref{fig:aerialhydro}.
Importantly, the detection functions defined in (\ref{eq:ds}) and (\ref{eq:am}) remain the same for each scenario.

\FloatBarrier

Next, we consider multiple levels of surface probabilities, $\pi$, and average number of calls per whale, $c$. For $\pi$, we consider values of 0.15 and 0.65 to be the low and high values, comparable to moderate, which was set to 0.40. For calls, $c$, we assume low and high values of 3 and 12, respectively, compared to the moderate value of 6. Again, both moderate specifications were used previously.

\begin{figure}
    \centering
\includegraphics[scale=.55]{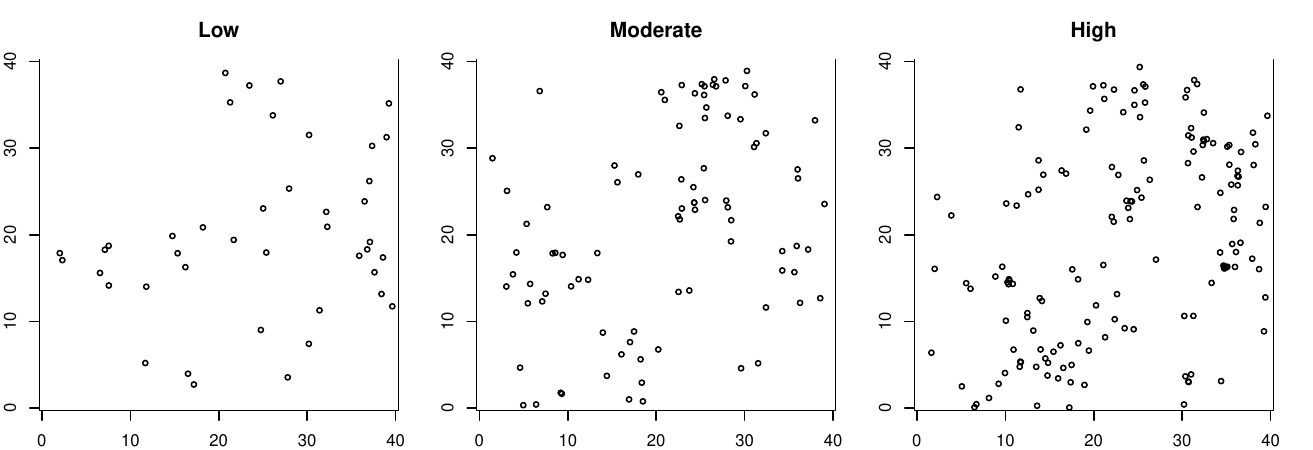}
    \caption{Simulated true whale locations based on low, moderate, and high abundance.}
    \label{fig:whales}
\end{figure}

Lastly, we vary the value of $\beta_0$ in order to simulate populations with low, moderate, and high abundance. The true intensity surface for moderate abundance is shown in Figure \ref{fig:True} (left) and was used in the simulation in Section \ref{sec3}.
The intensity surfaces for low and high abundance (not shown) are approximately 0.5 $\times$ and 2 $\times$ the intensity for moderate abundance, respectively. In our simulation, the low abundance setting resulted in 42 whales, moderate had 89 whales, and high had 152 whales. The locations of these simulated realizations are shown in Figure \ref{fig:whales}.

Table \ref{tab:mods} defines the 15 models that were considered for our simulation study based on varying combinations of aerial transect sampling intensity, PAM sampling intensity, surface probabilities, average number of calls per whale, and total abundance. These models are broken up into three groups: 1-9, 10-13, and 14-15. For the first group, we keep everything set to moderate except the aerial transect sampling intensity and PAM sampling intensity. This allows us to quantify the impacts of increasing and decreasing one or more of the data sources on estimating total abundance.
The second group assumes moderate sampling intensity of both sources while increasing and decreasing both the surface probability parameter, $\pi$, and the average number of calls per whale, $c$. One would assume that higher surface probabilities would result in better estimates of abundance from the aerial transect data. In addition, higher (lower) average call rates might increase (decrease) the utility of the PAM data.
The last group of models fixes all sampling intensities, surface probabilities, and average number of calls to their moderate values and investigates abundance estimation when total abundance is low versus high.
The low, moderate, and high values considered for each of the variables in the simulation are given in Table \ref{tab:LMH}.

\begin{table}
\caption{Simulation scenarios considered based on varying combinations of total abundance, aerial transect sampling intensity, PAM sampling intensity, surface probabilities, and average number of calls per whale.\label{tab:mods} }
\centering
\begin{tabular}{cccccc}
  \hline
        & Total     & Aerial & PAM          & Surface & Number  \\
Simulation   & Abundance & Intensity & Intensity & Probability &of Calls \\
  \hline
1   &   M  &   L    &   L    &    M   &    M   \\
2   &   M  &   M    &   L    &    M   &    M   \\
3   &   M  &   H    &   L    &    M   &    M   \\
4   &   M  &   L    &   M    &   M    &    M   \\
5   &   M  &   M    &   M    &   M    &    M   \\
6   &   M  &   H    &   M    &   M    &    M   \\
7   &   M  &   L    &   H    &   M    &    M   \\
8   &  M   &   M    &   H    &   M    &    M   \\
9   &  M   &   H    &   H    &   M    &     M  \\
\hline
10   &  M   &   M    &   M    &  L     &   L    \\
11   &  M   &   M    &   M    &   L    &   H    \\
12   &  M   &   M    &   M    &   H    &   L    \\
13   &  M   &   M    &   M    &   H    &   H    \\
\hline
14   &  L   &   M    &   M    &   M    &   M    \\
15   &  H   &   M    &   M    &   M    &   M    \\
   \hline
\end{tabular}
\end{table}

\begin{table}
\caption{The values of low, moderate, and high for each of the variables considered in the simulation. \label{tab:LMH}}
\centering
\begin{tabular}{lccccc}
  \hline
             & Aerial & PAM          & Surface & Number  & Total\\
Model    & Intensity & Intensity & Probability &of Calls & Abundance\\
  \hline
Low           &  4    &  4    &  $\pi=0.15$    &  $c=3$  &   42  \\
Moderate      &  8    &   9   & $\pi=0.40$     & $c=6$   &   89  \\
High        &  16    &   16   &  $\pi=0.65$    & $c=12$   &   152    \\
   \hline
\end{tabular}
\end{table}

\begin{figure}
    \centering
\includegraphics[scale=.55]{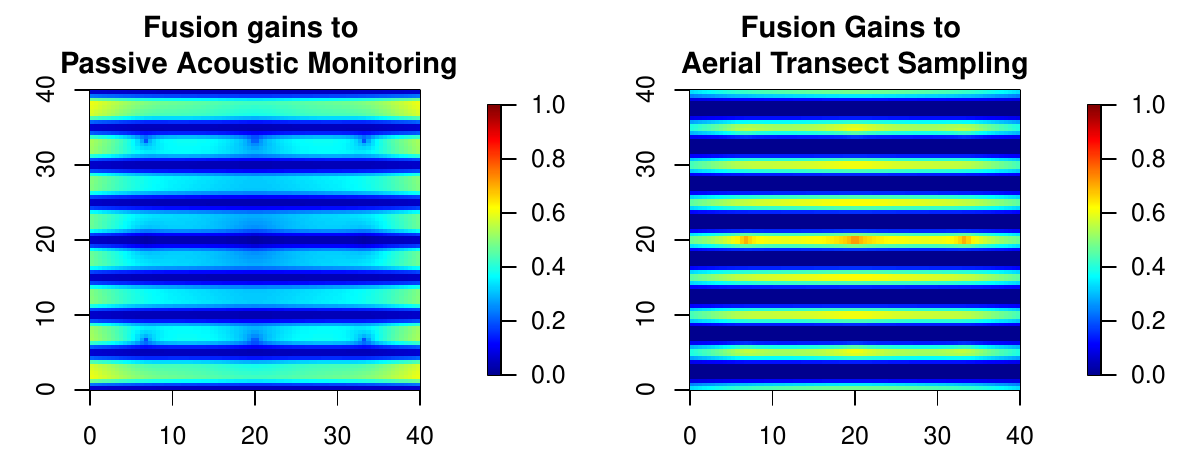}
    \caption{Difference in detection probabilities based on adding moderate aerial transect sampling intensity to moderate hydrophone sampling intensity (left) or adding moderate hydrophone sampling intensity to moderate aerial transect sampling intensity (right) given that a whale is on the surface and/or calling.}
    \label{fig:detection}
\end{figure}

We simulate realizations of the \emph{observed} aerial transect and PAM data following the same procedure outlined in Section \ref{sec3}.
That is, for each whale depicted in Figure \ref{fig:whales}, we randomly assign a binary variable for whether or not it is on the surface using a Bernoulli distribution with the specified surface probability, $\pi$. We also randomly simulate the number of calls made by each whale using the Poisson distribution with intensity equal to the average number of call value, $c$.
Then, the simulated realizations of the observed data are obtained based on the detection functions given in (\ref{eq:ds}) and (\ref{eq:am}). The resulting simulated datasets include locations of observed whales from the aerial transect sampling and the total number of calls recorded by each PAM.

For each simulation scenario listed in Table \ref{tab:mods}, we fit both single-source models as well as the proposed data fusion. MCMC is again used for model inference. The same prior distributions as outlined in Section \ref{sec2} are used and posterior inference is based on 15,000 posterior samples post 5,000 burn-in.

\begin{figure}
    \centering
\includegraphics[scale=.55]{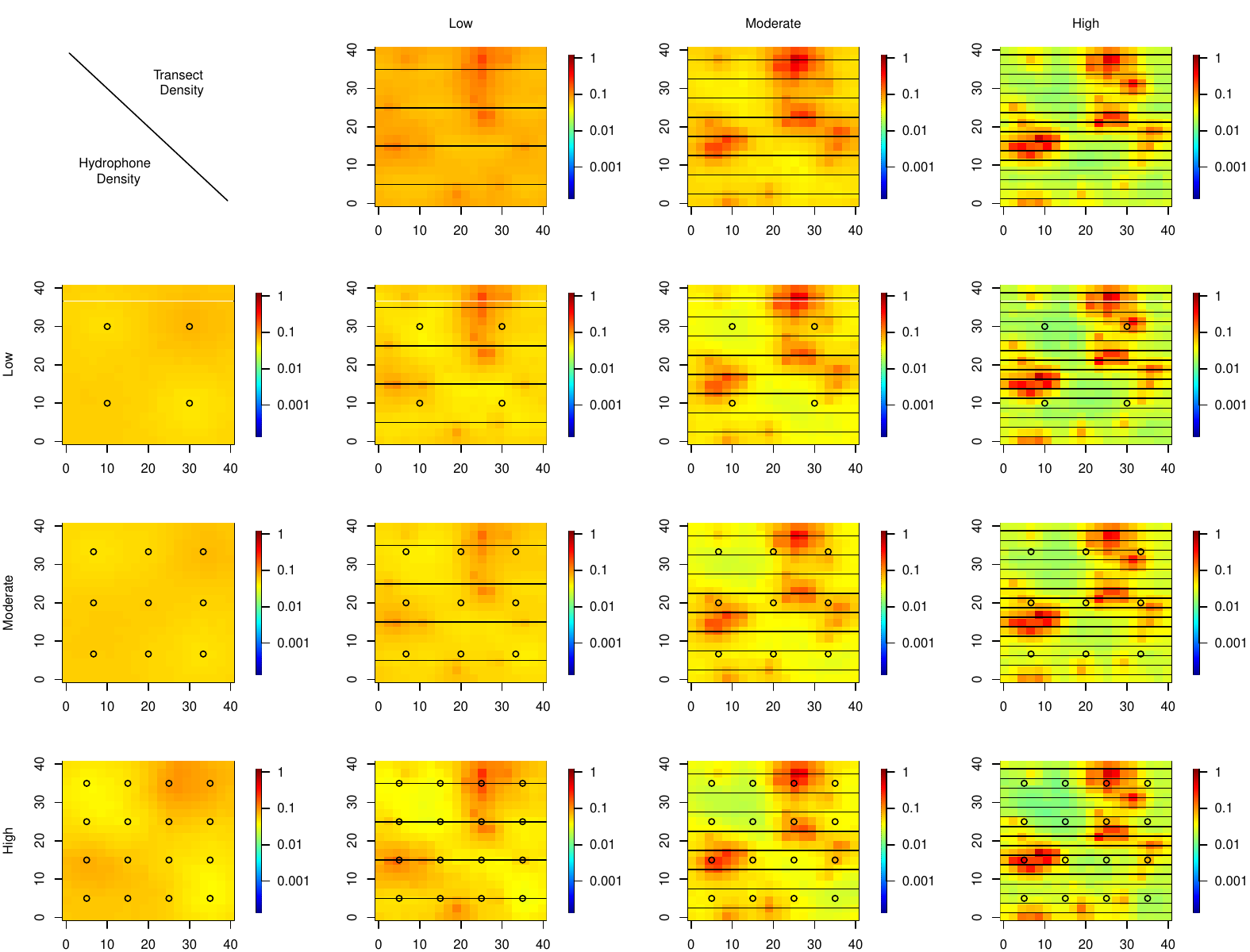}
    \caption{Posterior mean estimates of the intensity surface for the different combinations of aerial distance sampling and PAM sampling intensities. \label{fig:Est}}
\end{figure}

Figure \ref{fig:Est} shows the posterior mean surfaces of the intensity of abundance across the region for simulations 1-9 comparing low, moderate, and high sampling intensities. The first row and first column are the surfaces based on the models fitted using only a single data source. For the top row, we assume no PAM data and only low, moderate, and high aerial transect sampling intensities (see Figure \ref{fig:aerialhydro}, top). As sampling intensity increases, the resolution of the estimated intensity surface of abundance also increases. On the first column, only PAM data are used based on the low, moderate, and high intensity designs in Figure \ref{fig:aerialhydro} (bottom). Again, higher PAM sampling intensity results in more refined estimates of the intensity surface of abundance.

The results of the data fusion using low, moderate, and high sampling intensities of both data sources are shown in the lower right 3 $\times$ 3 panel of Figure \ref{fig:Est}. 
In general, model performance increases with increases in the sampling intensity of either or both data sources.
Higher PAM sampling or higher aerial transect sampling results in more refined estimates of the intensity surface, estimates of abundance, and lower RMSEs.
For a fixed aerial transect sampling intensity, the difference between the low, moderate, and high PAM sampling intensity is modest compared to the the difference between low, moderate, and high aerial transect sampling for a fixed PAM sampling intensity (Figure \ref{fig:Est}).

To further compare these fitted models given the simulated data, we also compute the posterior mean and standard deviation of the estimate of abundance, computed as the integral of the estimated intensity over the spatial domain. These estimates for each of the fitted models are given in Table \ref{table:A19}. Recall that the true number of whales in the simulation was 89, independent of the sampling intensities of the two data sources. Whereas each model fit captured the true value of abundance based on the 95\% credible interval, the estimates improved both in terms of accuracy and precision as the sampling intensity of either or both sources increased. In general, the PAM data appears to be slightly more valuable in terms of estimating total abundance, as evident by estimates close to the true value for most simulation scenarios.

Lastly, we compute the root mean squared error (RMSE) for these models and simulation scenarios using the discretized surface. Specifically, given the posterior mean of the intensity evaluated at each grid cell centroid, we compute the root mean square different between the log of the estimate and the log of the true intensity. Theses values are also reported in Table \ref{table:A19}. As expected, RMSE decreases as sampling intensity increases for either or both data sources. Again, high intensity sampling designs for either data source result in similar RMSE estimates and the data fusion outperforms the single source models.

\begin{table}
\caption{Posterior mean (standard deviation) estimates of total abundance given low, moderate, and high sampling intensity using hydrophones (rows) and aerial transects (columns). The first row and column correspond to the model fitted using only one data source. The true simulated abundance was 89. Also reported is the RMSE between the log of the posterior mean discretized intensity surface relative to the log of the true discretized intensity surface. \label{table:A19}}
\centering
\begin{tabular}{cccccc}
  \hline
        &&\multicolumn{4}{c}{Aerial Transect Intensity} \\
        && \multicolumn{1}{c}{-} & \multicolumn{1}{c}{Low} & \multicolumn{1}{c}{Moderate} & \multicolumn{1}{c}{High}  \\
  \hline
 &\multirow{2}{*}{-}                         &                   & 126.60 (36.82) & 118.63 (21.69) & 87.14 (13.03) \\
 &&                       & 2.10 & 1.90 & 1.59 \\
   &\multirow{2}{*}{Low}                  & 91.25 (6.74) & 92.88 (6.49) & 93.58 (6.17) & 88.94 (5.77) \\
PAM && 1.91 & 1.88 & 1.73 & 1.61 \\
Intensity      &\multirow{2}{*}{Moderate}    & 89.65 (4.55) & 90.68 (4.39) & 90.55 (4.29) & 87.77 (4.05) \\
 && 1.88 & 1.85 & 1.74 & 1.57 \\
 &\multirow{2}{*}{High}                      & 91.94 (3.93) & 92.69 (3.58) & 91.68 (3.40) & 90.26 (3.39) \\
& &1.81 & 1.82 & 1.67 & 1.56 \\
   \hline
\end{tabular}
\end{table}

Next we compare the simulation scenarios having varying surface probabilities and average number of calls (simulations 10-13). Table \ref{tab:A1013} gives the posterior mean and standard deviation estimates of total abundance and RMSE, respectively, for each simulation and model. Each model captures the true value of abundance based on the 95\% credible intervals. Uncertainty in the estimates decreases as the surface probability increases as well as when the average number of calls increases. The best model in terms of prediction accuracy and precision is the data fusion model with high surface probability and high average number of calls per whale. The RMSE estimates confirm these results.

\begin{table}[ht]
\caption{Posterior mean (standard deviation) estimates of total abundance given low and high numbers of calls (rows) and surface probabilities (columns).  The true simulated abundance was 89.\label{tab:A1013}}
\centering
\begin{tabular}{lc|ccc}
  \hline
   &&\multicolumn{3}{c}{Surface Probabilities} \\
    &    & - & Low & High \\
  \hline
&  \multirow{2}{*}{-}            &       & 	80.87  (28.57) & 108.23 (16.36) \\
 &&& 	 1.81 & 1.79 \\
\multirow{1}{*}{Number}&\multirow{2}{*}{Low}           & 96.27 (6.69) & 96.29 (6.22) & 96.10 (3.53) \\
\multirow{1}{*}{of Calls}& &  1.98 & 1.93 & 1.86 \\
&\multirow{2}{*}{High}                                   & 96.19 (3.80) & 97.49 (5.77) & 95.99 (3.21) \\
 &&  1.90 & 1.74 & 1.72 \\
   \hline
\end{tabular}
\end{table}

For the last simulations, we evaluate the fusion model for multiple levels of total abundance. In all previous simulations, we assumed moderate abundance in that the \emph{true} population consisted of 89 whales. Here, we assume low and high abundance where the \emph{true} population is simulated to be 42 and 152 whales, respectively. Under low and high abundance, each of the models captures the true value of the total population (Tables \ref{tab:A14}).
RMSE estimates for the low and high abundance simulations are also given in Table \ref{tab:A14}. Once again, the fusion model outperform both single data source models. In addition, the RMSE estimates are found to scale with total abundance. Recall in Table \ref{table:A19} that the RMSE estimate under the fusion model and moderate abundance was 1.74. Here, we see that the estimate is lower under low abundance and higher under high abundance.

\begin{table}[ht]
\caption{Posterior mean (standard deviation) estimates of total abundance given moderate sampling of aerial transects, PAM, and the fusion under low and high abundance. True abundance was 42 and 152. Also reported is root mean squared error between the log of the posterior mean discretized intensity surface relative to the log of the
true discretized intensity surface. \label{tab:A14}}
\centering
\begin{tabular}{l|cc}
  \hline
  & Low Abundance & High Abundance\\
  \hline
\multirow{2}{*}{Aerial Transect} & 45.54 (13.17) &135.16 (22.53)\\
&1.56 & 2.04\\
\multirow{2}{*}{Hydrophone}& 45.23 (3.13) &153.52 (6.69)\\
& 1.58 &2.19\\
\multirow{2}{*}{Fusion} & 44.57 (3.01) & 151.68 (5.85)\\
&1.51 &2.09\\
   \hline
\end{tabular}
\end{table}

\end{document}